\documentclass[tighten,preprint2,flushrt]{aastex}
\usepackage{geometry}
\usepackage{float}
\usepackage{epsfig,lscape} \usepackage[dvipdfm]{hyperref} \usepackage{natbib}
\usepackage{slashbox}
\usepackage{multirow}
\usepackage{subfigure}
\usepackage{color}
\usepackage{color,soul}
\usepackage{rotating}
\newcommand{\myemail}{zouhu@nao.cas.cn}

\begin{document}

\title{South Galactic Cap $u$-band Sky Survey  (SCUSS): Data Reduction}
\author{Hu Zou \altaffilmark{1}, Zhaoji Jiang \altaffilmark{1}, Xu Zhou \altaffilmark{1}, Zhenyu Wu \altaffilmark{1}, Jun Ma \altaffilmark{1}, Xiaohui Fan \altaffilmark{2}, Zhou Fan \altaffilmark{1}, Boliang He\altaffilmark{1}, Yipeng Jing \altaffilmark{3}, Michael Lesser \altaffilmark{2}, Cheng Li \altaffilmark{4}, Jundan Nie \altaffilmark{1}, Shiyin Shen \altaffilmark{4}, Jiali Wang \altaffilmark{1}, Tianmeng Zhang \altaffilmark{1}, Zhimin Zhou \altaffilmark{1}}
\altaffiltext{1}{Key Laboratory of Optical Astronomy, National Astronomical Observatories, Chinese Academy of Sciences, Beijing, 100012, China; \myemail}
\altaffiltext{2}{Steward Observatory, University of Arizona, Tucson, AZ 85721, USA}
\altaffiltext{3}{Center for Astronomy and Astrophysics, Department of Physics and Astronomy, Shanghai Jiao Tong University, Shanghai 200240, China}
\altaffiltext{4}{Shanghai Astronomical Observatory, Chinese Academy of Sciences, Shanghai 200030, China}

%%%%%%%%%%%%%%%%%%%%%%%%
%%%%%%%%%%%%%%%%%%%%%%%%
\begin{abstract} 
The South Galactic Cap $u$-band Sky Survey (SCUSS) is a deep $u$-band imaging survey 
in the Southern Galactic Cap, using the 90Prime wide-field imager on the 2.3 Bok telescope at Kitt Peak. The survey observations
started in 2010 and ended in 2013. The final survey area is about 5000 
deg$^2$ with a median 5$\sigma$ point source limiting magnitude of $\sim23.2$. This paper describes the survey data reduction process, which 
includes basic imaging processing, astrometric and photometric calibrations, 
image stacking, and photometric measurements.  Survey photometry is performed on 
objects detected both on SCUSS $u$-band images and in the SDSS database.  
Automatic, aperture, point-spread function (PSF), and 
model magnitudes are measured on stacked images. Co-added aperture, PSF, and model 
magnitudes are derived from measurements on single-epoch images. We also present comparisons of the SCUSS photometric catalog with those of the SDSS and Canada–France–Hawaii Telescope Legacy surveys. 
\end{abstract}


%%%%%%%%%%%%%%%%%%%%%%%%%
\section{Introduction}
The South Galactic Cap $u$-band Sky Survey (SCUSS; see X. Zhou et al. 2015, in preparation for more  
detail) is a deep imaging survey in the South Galactic Cap in $u$-band with an effective 
wavelength of 3538 \AA. It is an international cooperative project between the 
National Astronomical Observatories of China and the Steward Observatory of the University of Arizona. The 
survey utilizes the 2.3m Bok telescope at Kitt Peak. The camera, installed 
at the prime focus, provides a field of view (FOV) of about 1 deg$^2$. The 
adopted filter is similar to the $u$ band of the Sloan Digital Sky Survey 
\citep[SDSS;][]{yor00}. The SCUSS project started in the summer of 2009 and began its 
observations in fall 2010. Survey observations were completed in the fall of 2013. The 
final survey area is about 5000 deg$^2$ with uniform imaging depth,  
of which more than 75\% is covered by the SDSS footprint. The median imaging depth for point 
sources is about 23.2 mag at a signal-to-noise ratio (S/N) of 5 with a 5 minute 
exposure time. Table \ref{tab1} gives the basic characteristics of SCUSS. 

\begin{table}[!h]
\centering
\caption{Survey Summary\label{tab1}}
\begin{tabular}{ll}
\hline
\hline
Telescope & 2.3 m Bok telescope \\
Site & Kitt Peak in Arizona \\ 
CCD & 2$\times$2 4k$\times$4k CCD array \\
Field of view & 1\arcdeg.08$\times$1\arcdeg.03 \\
Filter & $u$ (3538\AA) \\
Integral time & 300 s \\
Magnitude limit & 23.2 mag \\
Survey area     & $\sim$5000 deg$^2$ \\
Observation period & 2010--2013 \\ 
\hline
\end{tabular}
\end{table}

The main goal of the survey is to supply input photometric catalogs to select spectroscopic targets for the 
Large Sky Area Multi-Object Fiber Spectroscopy Telescope \citep{cui12}. 
In addition, by combining with other bands in large-scale photometric surveys, such as the SDSS 
and Panoramic Survey Telescope \& Rapid Response System \citep[Pan-STARRS][]{kai04},  
the survey data can be used for a wide
range of scientific investigations, such as Galactic structure, 
Galactic extinction, galaxy photometric redshift, galaxy star formation rates, and stellar 
populations of nearby galaxies. 

This paper describes the data reduction pipeline specially designed 
for the SCUSS survey. There are some instrumental issues that need to be specially 
handled, such as issues in overscan, substructures in bias, and crosstalk.
In addition to detecting sources on SCUSS images, we also provide photometry for SDSS objects using consistent
object parameters. Section \ref{sec2} introduces the SCUSS 
survey and related facilities. Section \ref{sec3} describes the basic image 
processing. Astrometric and photometric calibrations are presented in Sections 
\ref{sec4} and \ref{sec5}, respectively. Section \ref{sec6} provides image-quality 
statistics of the observations. Section \ref{sec7} describes image stacking. 
Sections \ref{sec8} and \ref{sec9} present photometry methods and a  
comparison with other surveys, respectively. A summary is given in Section \ref{sec10}.

%%%%%%%%%%%%%%%%%%%%%%%%%%%%%%%%%%%%%%%%%%%%%%%%%%%%%%%%%%%%%%%%%%%%%%%%%%%%%%%%%%%%%%%%%%%%%%%%%%%%%%%%%%%%%
\section{The survey and facilities} \label{sec2}
SCUSS is a $u$-band imaging survey in the northern part of the southern Galactic 
Cap. The survey originally covered the region of Galactic latitude $b < -30{\arcdeg}$ and 
equatorial latitude $\delta > -10\arcdeg$, with a total area of about 3700 deg$^2$. 
It was further extended to the Galactic anti-center region and the extra area covered 
by the SDSS, with a final survey area of about 5000 deg$^2$ (X. Zhou et al. 2015, in preparation). 
Normally, there are two exposures for each field and the total exposure time is 5 minutes, which generates 
 images  1--1.5 mag deeper than the SDSS $u$ band. In the following, we give a general 
description of the telescope, camera, and filter that were used in SCUSS. 

\subsection{Telescope}
The Bok telescope \footnote{http://james.as.arizona.edu/~psmith/90inch/90inch.html} 
is a 90 inch (2.3 m) telescope operated by the Steward Observatory of the University of Arizona. It is located 
on Kitt Peak, whose latitude is +30\arcdeg57\arcmin46\arcsec.5 and longitude is
111\arcdeg36\arcmin01\arcsec.6W. The elevation is about 2071 m and the typical 
seeing is  about 1\arcsec.5. The telescope runs year-round except 
on Christmas Eve and during a maintenance period in August. The accuracy of the 
absolute pointing of the telescope is recorded as 3{\arcsec} over the entire sky. 

\subsection{Camera}
\begin{figure}[!h] \epsscale{1.0} 
\plotone{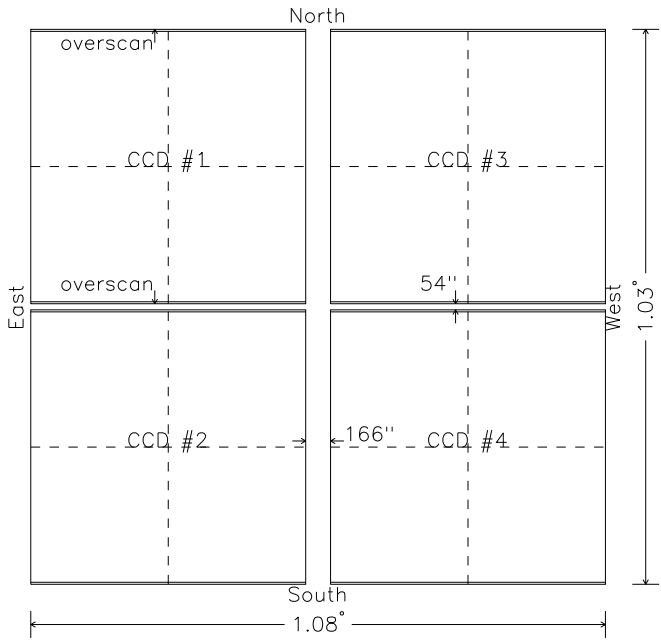} \caption{Layout of the CCD array. There are four CCDs: 
CCD \#1, \#2, \#3, and \#4. Each CCD has four amplifiers highlighted by 
dashed lines. Overscan regions are also indicated. 
\label{fig1}} 
\end{figure}

An imaging system, named 90Prime, is deployed at the prime focus (corrected 
focal ratio: $f$/2.98; corrected focal length: 6829.2 mm). The detector is a 
CCD array consisting of four 4k$\times$4k backside-illuminated CCDs. They are 
STA2900 CCDs made by Semiconductor Technology Associates, Inc. and backside 
processed at the University of Arizona Imaging Technology Laboratory.  These CCDs 
have been optimized for the $u$-band response, giving a quantum efficiency at $u$ 
band of about 80\%. Figure \ref{fig1} displays the layout of the CCDs on the focal plane. The edge-to-edge FOV is about 1\arcdeg.08$\times$1\arcdeg.03. The pixel 
scale is 0\arcsec.454. There are inter-CCD spacings in the center of the array: 
166{\arcsec} in right ascension and 54{\arcsec} in declination.  Each CCD is read out 
by 4 amplifiers located at the corners. Each detector has 4096$\times$4032 physical pixels and 
20-row pixels of overscan for each amplifier. The full well is about 90,000 electrons or 
65,000 data numbers (DNs). The current system gain is set to be about 1.5 electrons 
per DN. The dark current is about 7 electrons per pixel per hr. The readout time is 
about 30 s and the average readout noise is about 8.8 electrons. The 
response non-uniformity with the $u$ filter, which is the standard deviation divided 
by the mean of a $u$ band flat field image, is about 2\%. Table \ref{tab2} summarizes 
these parameters of the detector. In 2010, CCD \#4 had problems and only a quarter of 
this CCD could be used. CCD \#2 and CCD \#3 had relatively large readout noise. 
The camera was upgraded in 2011, at which time these two CCDs were replaced with new 
detectors. CCD \# 4 was replaced and swapped with CCD \#1. New video 
preamps were added to all four CCDs in order to reduce crosstalk and improve noise immunity.

\begin{table}[!h]
\small
\centering
\caption{Detector Parameters\label{tab2}}
\begin{tabular}{ll}
\hline
\hline
CCD number & 4 (in 2$\times$2 array) \\
CCD gaps & 166{\arcsec} in R.A. and 54{\arcsec} in decl. \\
CCD size   & 4096$\times$4032 (extra 40 rows of overscan) \\
Full well  & 90,000 electrons/pixel (65,000 ADUs/pixel) \\
Amplifier number & 4 \\
Quantum efficiency & $\sim$80\% at 3500\AA \\
Average dark current & 7.2 electrons/pixel/hr \\
Response non-uniformity & 2\% at $u$ band\\
Average readout noise & 8.8 electrons \\
Readout time & $\sim$30 s \\
Average gain & 1.5 electrons/DN \\
Plate Scale & 0\arcsec.454 (15 $\mu$m) \\
\hline
\end{tabular}
\end{table}

\subsection{Filter}\label{sec-filt}
\begin{figure}[!h] \epsscale{1.0} 
\plotone{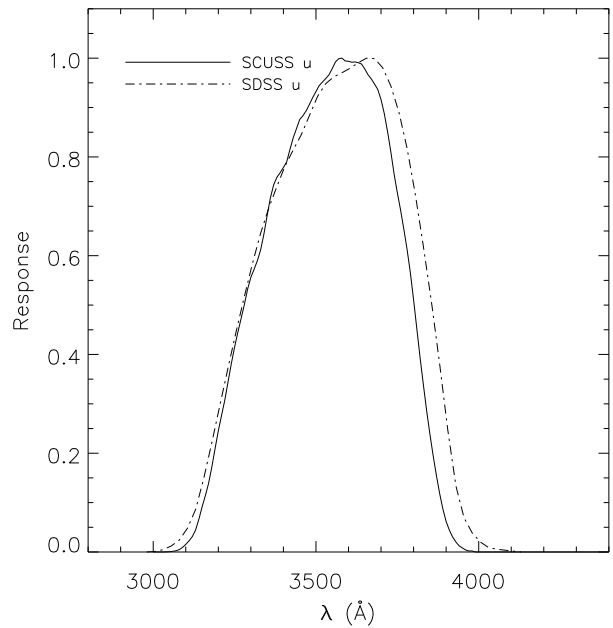} \caption{Response curves of the SCUSS $u^*$ and SDSS $u$ 
filters. Both curves include the atmospheric extinction at an airmass of 1.3 
and normalized to their maximums. \label{fig2}} 
\end{figure}

The SCUSS $u$ filter is similar to the SDSS $u$ band. Figure \ref{fig2} 
displays both SCUSS and SDSS system response functions. The SCUSS $u$ 
response curve includes the filter transmission, the CCD quantum efficiency, 
and the atmospheric extinction at the typical airmass of 1.3.
The adopted atmospheric extinction is the same as that of the SDSS, which
is based on the standard Palomar monochromatic extinction coefficients but 
here scaled to the elevation of Kitt Peak assuming an exponential scale height 
of the atmosphere of 7000 m \citep{doi10}. The effective wavelength and bandwidth of 
the filter are respectively defined as $\lambda_\mathrm{eff}=\frac{\int{\lambda}R(\lambda)d\lambda}{{\int}R(\lambda)d\lambda}$ and $\lambda_\mathrm{eff}=\frac{\int{\lambda}R(\lambda)d\lambda}{{\int}R(\lambda)d\lambda}$, 
where $R(\lambda)$ is the filter response curve. The effective wavelength and bandwidth 
of the SCUSS $u$ band are about 3538 {\AA} and 345 {\AA}, respectively. The FWHM is 
about 520 \AA. The effective wavelength, bandwidth, and FWHM of the 
SDSS $u$ filter is 3562, 385, and 575 \AA, respectively. The SCUSS $u$ band 
is slightly bluer than the SDSS filter. In the rest of this paper, we will 
use the symbol of $u^*$ to refer to the SCUSS $u$ band and the symbols of $u$, 
$g$, $r$, $i$, and $z$ for the five SDSS bands. Throughout the paper, objects are classified as 
point-like or extended throughout this paper using the SDSS star--galaxy separation unless otherwise specified.

\section{Basic image processing}\label{sec3}
\subsection{Image division and overscan correction}
90Prime has four CCDs and each CCD is read out by four amplifiers. 
Thus, every exposure file includes 16 FITS extensions. We split 
the file into four smaller FITS images, each of which represents one 
of the four CCDs. Overscan lines (40 pixels) are moved to the right 
side of the image.

We compute the median of the overscan columns for each amplifier and 
subtract it from the raw frame. There are some subtle issues about 
the overscan that require special care. For example, when a bright star is located right beside 
the overscan region, nearby overscan pixels are contaminated. 
Occasionally, some brighter stripes appear within or close to either 
end of the overscan region. The overscan in these regions needs to be 
interpolated or extrapolated. After overscan subtraction, frames are 
rotated counter-clockwise by 90{\arcdeg} to keep north up and east 
left. These frames are then trimmed to the size of 4096$\times$4032 pixels. 

\subsection{Dark Current}
The average dark current of the 90Prime imager is about 7.2e per 
hr. For an exposure of 300 s, it is ignorable (about 0.6e) relative 
to the readout noise of about 8.8e, so we did not take dark exposures.

\subsection{Bias}
\begin{figure}[!h]
\centering
\plotone{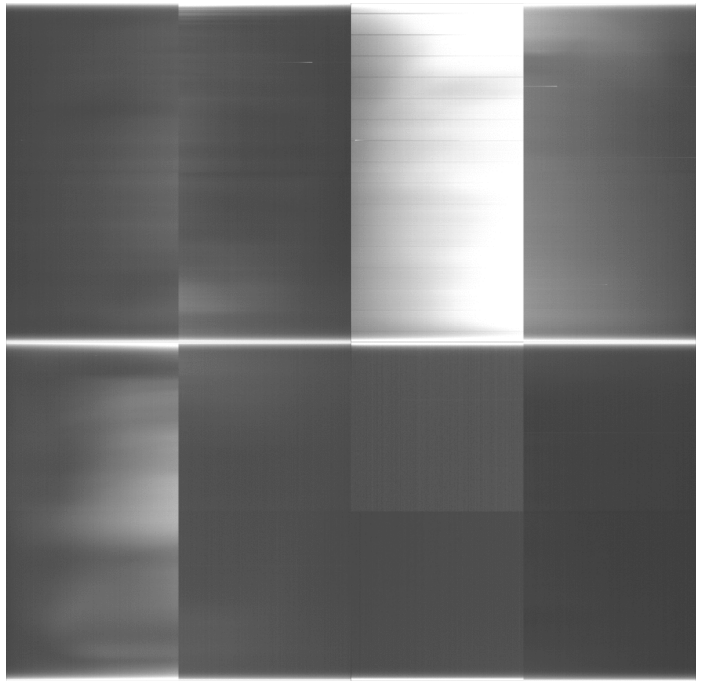}
\caption{"Super" bias image of four CCDs in 2011 December. The 
layout of these CCDs is the same as that shown in Figure \ref{fig1}. 
\label{fig3}}
\end{figure}

Bias frames are taken before and after the scientific exposures. 
A total of 20 bias frames are obtained each night. After combining them, 
there are large-scale structures, especially for CCD 
\#2 and \#3 as shown in Figure \ref{fig3}. Counts in some peaks of 
the structures can be more than 10 DN. These structures affect 
the accuracy of photometry significantly, especially in $u$ band, because 
counts of scientific frames in this band are low. The bias 
structures are stable for a long time (month-to-month variation is 
about 0.25 ADU), so we derive a median "super" bias by combining all 
the bias frames taken within a month. All overscan-subtracted frames 
are corrected by this "super" bias. 

\subsection{Flat-fielding}
Flat-fielding removes the instrumental signatures from raw 
frames using several exposures taken with the telescope facing a uniform 
light source. At the beginning and end of each night, a total of 20 dome 
flats are taken with the dome closed and the telescope pointing to a white 
screen. This screen is illuminated by UV lamps of Philips MasterColor 
Ceramic Metal Halide ED-17. Usually, a 6 s exposure generates 
about 20,000 DN on the CCD. Although dome flats have extremely high 
S/Ns, there are some disadvantages to applying them in flat-fielding. 
First, lamps are point sources and the scattered light on the 
screen might be uneven in the radial direction. Second, the optical path of the light from the screen 
is not the same as the light from night sky. In addition, we also 
find that the gain of each amplifier changes considerably during the 
whole night. When using only dome flats, we find that average 
sky backgrounds in the four amplifiers of each CCD can be quite different. 

Twilight flats are an alternative way and regarded as being more uniform 
than dome flats. They are obtained by observing the sky during evening 
and morning twilight. The brightness of the twilight sky changes 
rapidly and it strongly depends on the weather. It is difficult to 
get enough good high S/N flats. The problem of gain variation also 
persists.

\begin{figure}[!h]
\centering
\plotone{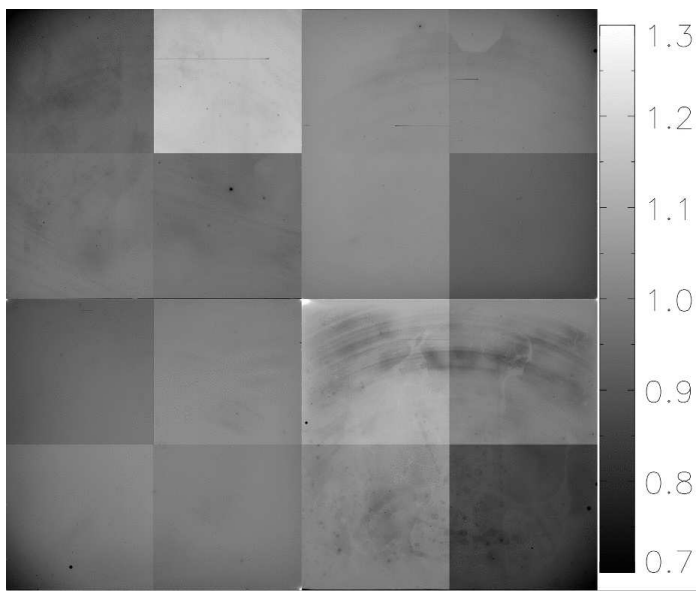}
\caption{"Super" sky flat image of four CCDs on 2011 December 28. The 
layout of these CCDs is the same as that shown in Figure \ref{fig1}. 
\label{fig4}}
\end{figure}

Raw science images also contain the flat-field characteristics. By combining 
all the science frames with outliers rejected, we can obtain a "super" 
sky flat. This kind of flat has more advantages than the  two 
types of flats above. The light of night sky "super" flat comes through the same optical path
as the observed objects. The CCD gain is the average of all science frames and 
it is closer to the real-time variation during the night. There are  
some structures in the "super" flat that present a different sensitivity 
amplitude relative to the dome flats possibly due to a different light path.
About 150 science images on average are observed each night. The 
average level of the sky background in each image is about 150 ADU. 
As a result, the "super" sky flat has an average count of about 23,000 ADU.
We present the sky flats on 2011 December 28 as an example in 
Figure \ref{fig4}. After applying either the dome flat (higher S/N) 
or super sky flat, we find that there is less than a 0.2\% of difference 
in the sky background fluctuations. The resulting error due to lower S/N 
of the "super" flat is negligible  relative to the large sky background 
fluctuation of about 12.7 ADU and CCD readout noise of 8.8 electrons. 
Thus, we use the "super" sky flat to do flat-fielding corrections 
of the science frames. 

\subsection{Crosstalk}
Crosstalk frequently occurs in multi-channel CCD chip 
read-out. When there are saturated objects in one of the CCD amplifiers, 
it can be best seen as mirror images in other amplifiers. The crosstalk 
usually causes contaminations across the output amplifiers at the level 
of 1:10,000 \citep{fre01}. The output counts of each amplifier are the 
sum of the true counts from the sky and a small fraction of counts from 
other amplifiers. The crosstalk signal can be either positive or negative 
and should be corrected for high-precision astronomical photometry. 

\begin{figure*}[!htbp] \epsscale{1.0}
\centering
\plotone{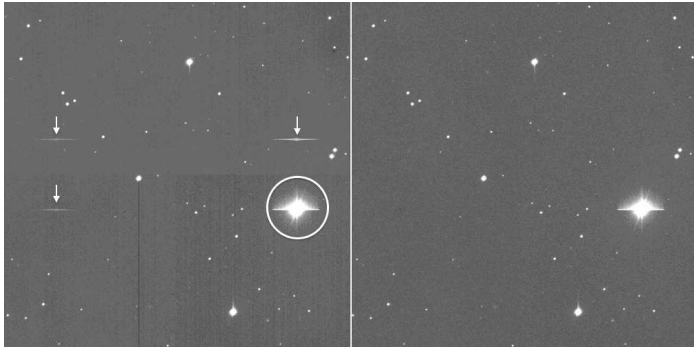}
\caption{Left: a SCUSS raw CCD image with visible crosstalk signals. 
The arrows point to the mirror images of the bright star marked with a circle. 
Right: the image after correcting the crosstalk. \label{fig5}}
\end{figure*}

The 90Prime camera has four CCDs and each CCD has four amplifiers. The 
crosstalk effect is clearly seen in our SCUSS raw images with saturated 
stars. The crosstalk signal is positive. We characterize the effect  assuming that it is additive and proportional 
to the number counts of other amplifiers. The proportionality coefficients are 
relatively stable in SCUSS images. We correct crosstalk using the following prescription:
 a series of images with bright stars appearing in one of the CCD 
amplifiers is selected; the proportionality coefficients are 
estimated by comparing the mirror signals with their original signals; the 
crosstalk signals in one quadrant are removed with the corresponding 
coefficients of the other three quadrants. The average coefficient for the 
SCUSS images is about 2:10,000. Figure \ref{fig5} illustrates the crosstalk 
and the performance of its correction. The crosstalk signals are clearly 
seen as mirror images (arrows) of a bright star that is marked with a 
circle. 

\section{Astrometric calibration} \label{sec4}
\subsection{Astrometry by UCAC4}
Astrometric solutions are derived by cross-identifying objects in science 
frames with the Fourth US Naval Observatory CCD Astrograph Catalog 
\citep[UCAC4]{zac13}. UCAC4 is an all-sky star catalog that 
is complete in $R$ band down to 16 mag. It contains proper motions 
for most stars. An approximate first-order guess of the astrometric 
solution for each CCD is first estimated by using UCAC4 objects in the 
same area with positions revised according to their proper motions. 
Then, the projection center of the telescope can be calculated using  
this rough solution. A more accurate astrometric solution is then  
derived with the projection center and a radial second-order correction 
for the focal plane distortion. This distortion term is the same for all frames 
in our survey.  

\begin{figure*}[!htbp] \epsscale{1.0}
\subfigure{\includegraphics[width=0.5\textwidth]{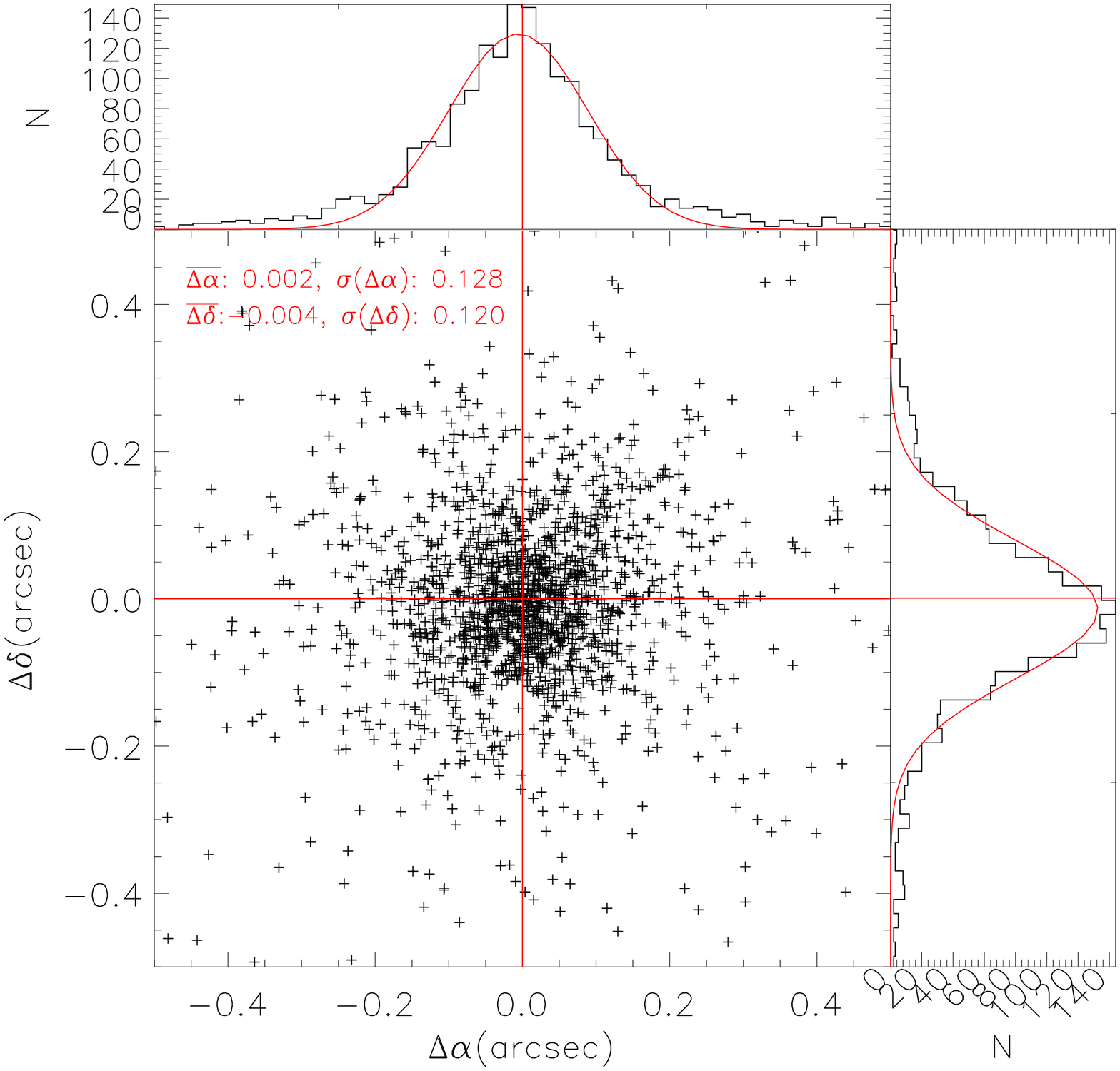}}
\subfigure{\includegraphics[width=0.5\textwidth]{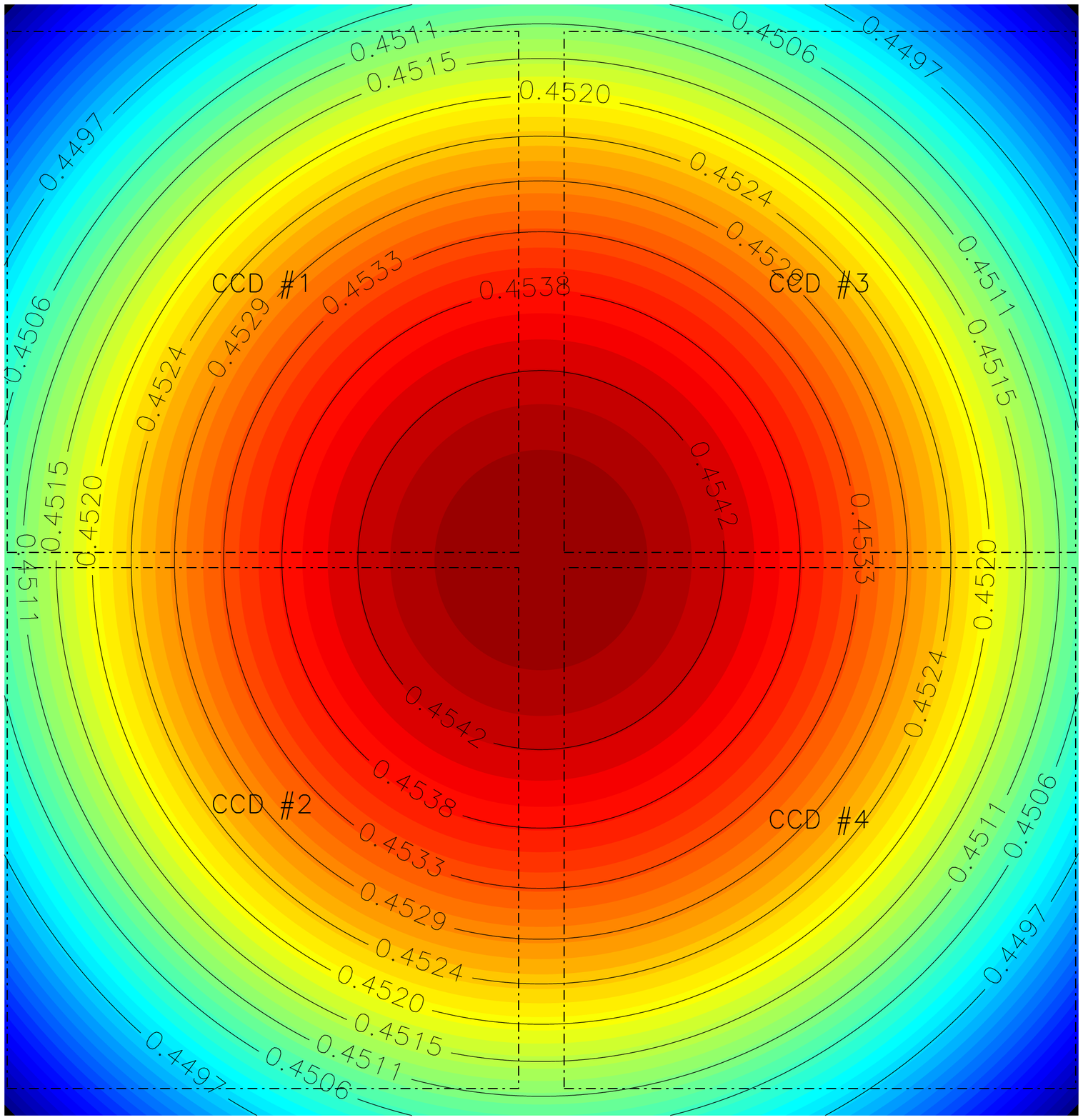}}
\caption{Left: distribution of astrometric errors in the plane of R.A. and 
decl. differences between SCUSS and UCAC4 positions for a randomly selected field. 
The red cross denotes the coordinate origin. Histograms showing the marginal 
distributions of the R.A. and decl. errors are plotted in the upper and 
right panels, where the red curves show the Gaussian fits. The means 
and standard deviations of the two distributions are also presented in 
the plot. Right: pixel scale variation in the detector plane. The layout of 
the CCD array (dashed boxes) is also plotted. Numbers labeling the 
contours give the pixel scales in arcseconds, which are also indicated by 
the colored surface.
\label{fig6}}
\end{figure*}

The left panel in Figure \ref{fig6} shows the distribution of the 
astrometric errors in the plane of R.A. and decl. differences between 
SCUSS and UCAC4. It includes objects in one exposure of a randomly 
selected field. About 1800 objects are matched with UCAC4. The average 
position offsets are 0\arcsec.002 and 0\arcsec.004 for R.A and decl., 
respectively. The 1$\sigma$ position errors in R.A. and decl. are about 
0\arcsec.128 and 0\arcsec.120. The right plot of Figure \ref{fig6} 
illustrates the distortion of the focal plane. The plate scale varies 
from the center (0\arcsec.455) to the edge (0\arcsec.447) of the FOV. 
The pixel area near the corner is about 3.3\% smaller than that of the 
central pixel.

\subsection{External astrometric errors}
External astrometric errors are estimated by matching SCUSS objects 
with the UCAC4 catalog. About 165 objects on average for each CCD 
are crossing-identified and the global external position error is about 
0\arcsec.13$\pm$0.02. The mean external astrometric offsets and RMS  
errors for R.A. and decl. are
\begin{eqnarray}
\nonumber
\overline{\Delta{\alpha}} = -0\arcsec.0012\pm0.0079, \overline{\Delta{\delta}} = -0.\arcsec0015\pm0.0071, \\
\nonumber
\sigma_{\Delta{\alpha}} = 0\arcsec.1111\pm{0.0178}, \sigma_{\Delta{\delta}} = 0\arcsec.1094\pm{0.0182},
\end{eqnarray}
where $\sigma$ corresponds to the 68.3\% confidence level.

\begin{figure*}[!htbp] \epsscale{1.0}
\subfigure{\includegraphics[width=0.5\textwidth]{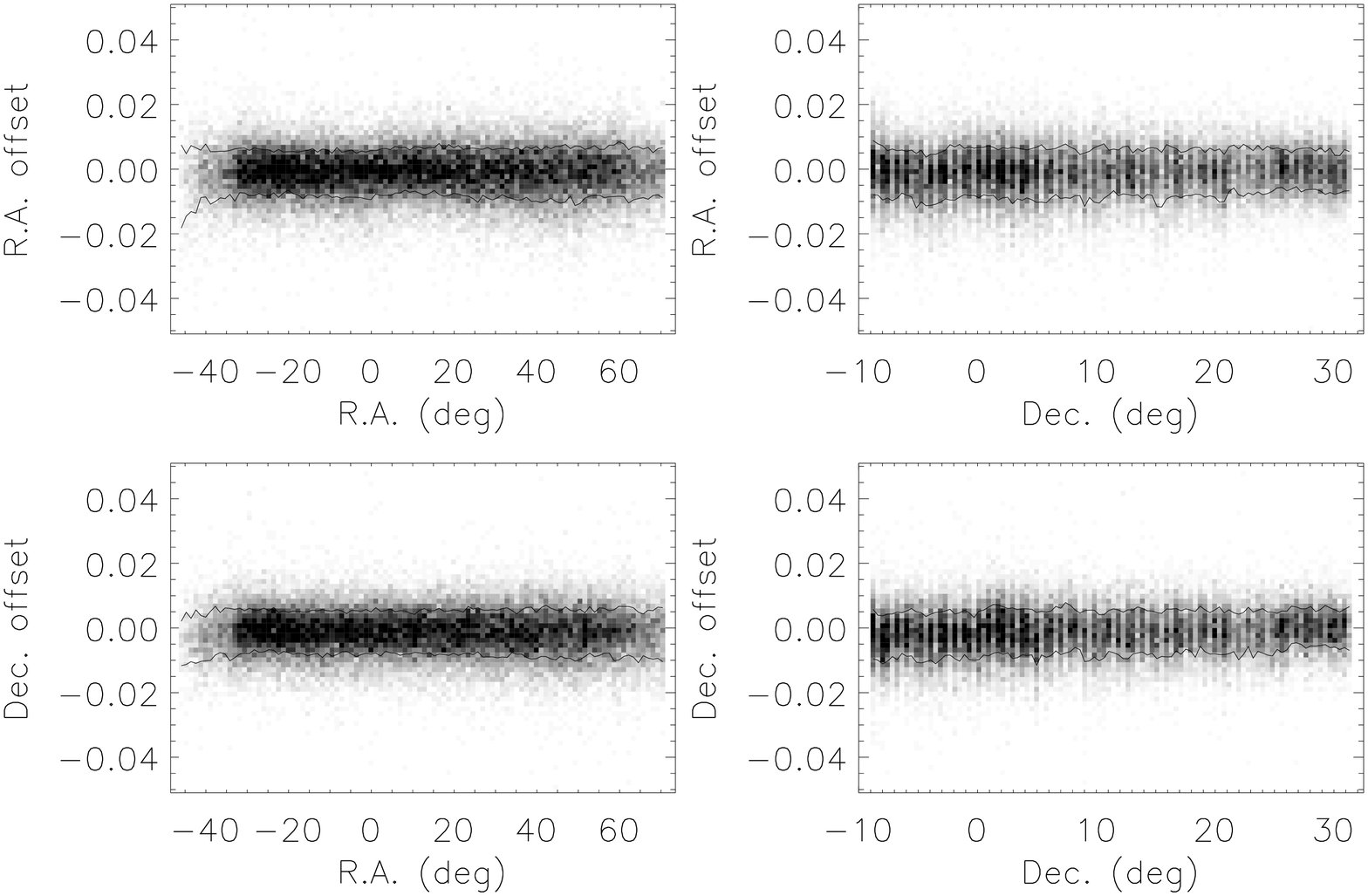}}
\subfigure{\includegraphics[width=0.5\textwidth]{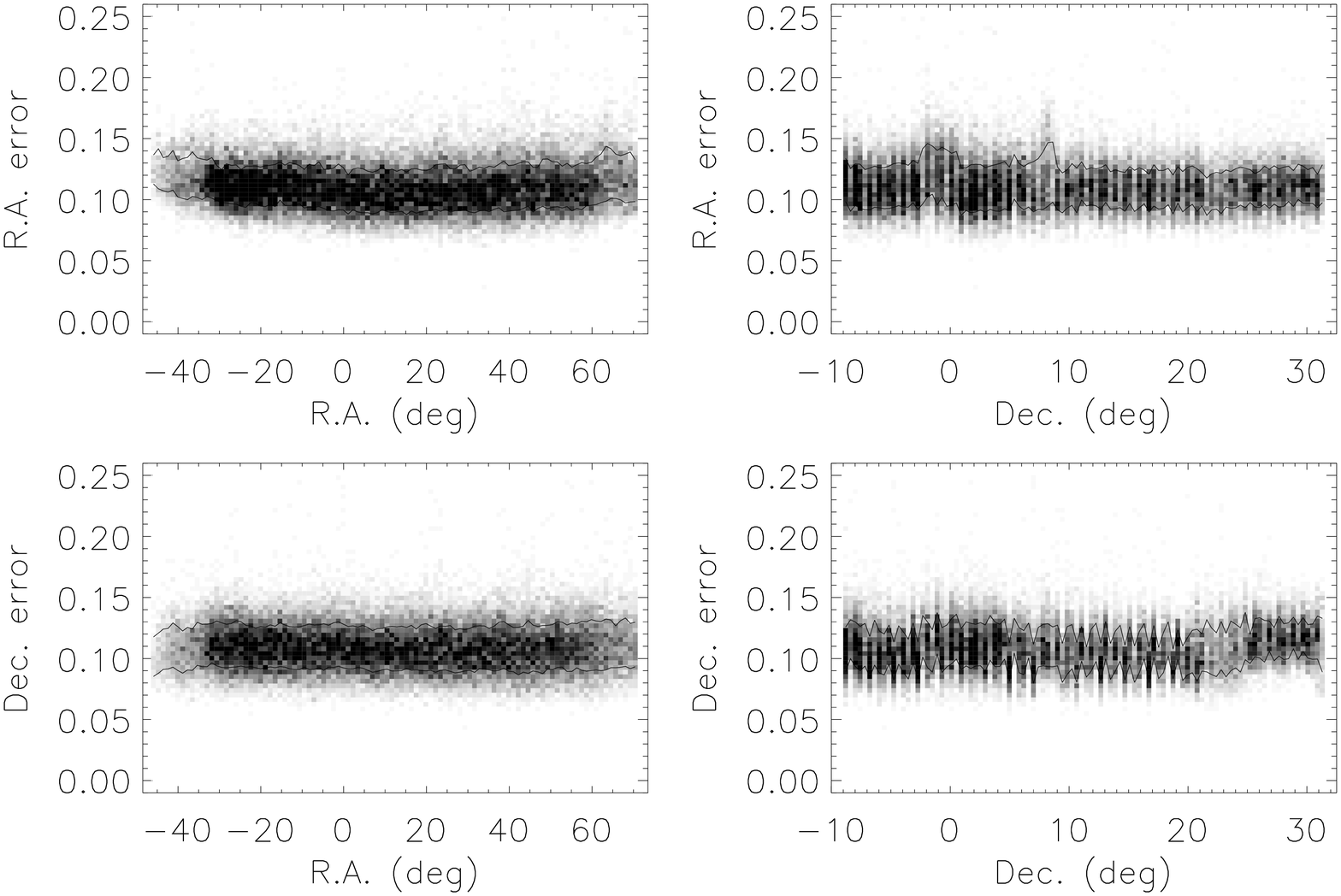}}
\caption{SCUSS R.A. and decl. offsets and RMS errors relative to UCAC4 
as functions of the equatorial longitude and latitude. The two curves 
in each panel show $\pm1\sigma$ scatter around the average along the 
$x$ axis. The R.A. is transformed to the range of (-180\arcdeg,180\arcdeg).
\label{fig7}}
\end{figure*}

Figure \ref{fig7} displays the external coordinate offset 
and RMS error as functions of RA and decl. 
The offsets and RMS errors are uniform over the entire survey. 

\subsection{Internal astrometric errors}
Internal astrometric calibration errors are estimated by using the 
cross-identifications of UCAC4 sources inside the overlapping areas 
between two adjacent exposures. Each field is observed with two 
exposures dithered by half a CCD size, so every CCD frame is 
covered by four other frames, whose astrometric solutions are 
independently derived. We calculate the internal astrometric 
uncertainties using the objects in common to both frames. The global 
average internal error is about 0\arcsec.09$\pm$0.03. The mean 
internal R.A. and decl. offsets and RMS errors over the whole survey 
are 
\begin{eqnarray}
\nonumber
\overline{\Delta{\alpha}} = 0\arcsec.0009\pm0.0102, \overline{\Delta{\delta}} = 0.\arcsec0004\pm0.0113, \\
\nonumber
\sigma_{\Delta{\alpha}} = 0\arcsec.0728\pm{0.0295}, \sigma_{\Delta{\delta}} = 0\arcsec.0707\pm{0.0269}.
\end{eqnarray}

\section{Photometric Calibration}\label{sec5}
\subsection{External calibration}
The SDSS imaging survey has covered more than one third of the sky within 
both northern and southern Galactic caps. It provides 
photometric catalogs of about 5200 square degrees in the SGC. More than 
21 million objects in this area are recorded. The exposure time for 
each band is about 54 s. The $u$ magnitude limit with 95\% 
completeness for point sources is about 22.0 mag. The photometric 
calibration accuracy in $u$ band is about 1.3\%. The ninth 
data release \citep[DR9]{ahn12} is utilized to make photometric 
calibrations of the SCUSS images.

\begin{figure}[!h] \epsscale{1.0}
\plotone{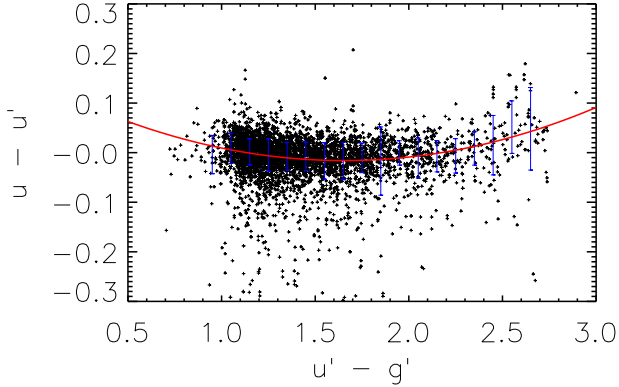}
\caption{Magnitude difference between the SCUSS $u^*$ and SDSS $u$ 
as a function of the SDSS $u - g$ color. The points are for point-like 
objects with $16 < u < 20.5$. The error bars show the averages and 
1$\sigma$ errors within different color bins. The red curve is the fitted transformation 
equation as shown in Equation (\ref{equ3}). \label{fig8}}
\end{figure}

Because of gain and observational condition variations, individual 
zeropoints must be determined separately for each frame. Aperture photometry 
is performed by using DAOPHOT \citep{ste87} on the bias-subtracted and flat-fielded 
images. We choose an aperture radius of about 7\arcsec.3 (16 pixels), similar 
to that of 7\arcsec.43 adopted by the SDSS photometric calibration. The aperture 
diameter is about 7 times the typical seeing FWHM, which is large enough not to be affected by 
aperture correction. The transformation equation between the instrumental 
magnitudes $u_\mathrm{inst} = -2.5\mathrm{log_{10} ADU}$ to the SDSS 
calibrated magnitude is simply given by 
\begin{equation}
u = u_\mathrm{inst} + c + kX + f(u - g) + f^\prime(u - g,X), \label{equ1}
\end{equation}
where $c$ is the instrumental zeropoint, $k$ is the atmospheric extinction 
coefficient, $X$ is the airmass, $u - g$ is the SDSS color, 
and $f(u - g)$ is the color term. The final term of $f^\prime(u - g,X)$ is related to the 
color effect due to the atmospheric extinction. The central wavelength of SCUSS $u^*$ 
band at an airmass of 1.0, 1.3, and 1.5 is estimated to be 3534\AA, 3538\AA, and 
3542\AA. The photometric effect in the color terms for most main-sequence stars 
at different airmasses is less than 0.3\%, estimated using theoretical 
stellar spectral libraries. So, we ignore the second-order color term. 
Since the standard stars are the common SDSS objects within the same area of each image, 
the term solely related to the airmass is constant. 
Thus, Equation (\ref{equ1}) can be simplified as 
\begin{equation}
u^* = u - f(u - g) = u_\mathrm{inst} + C, \label{equ2}
\end{equation}
where $u$ is the calibrated SCUSS $u$-band magnitude and $C$ is still termed 
as the "photometric zeropoint".  The color term is estimated as
follows: (1) SCUSS instrumental magnitudes are calibrated using the photometric 
magnitudes of common SDSS objects without considering the color term; (2) magnitude 
differences between these calibrated SCUSS magnitudes and SDSS magnitudes are 
calculated as a function of the SDSS $u - g$ color; (3) an 
approximate system transformation formula is fitted by a two-order polynomial:  
\begin{eqnarray}
\nonumber
u^* - u &=& - f(u - g) \\ 
\nonumber
             &=& 0.1459 - 0.1957(u - g) \\
             & & + 0.0591(u - g)^2, \label{equ3}                       
\end{eqnarray}
for  $0.8 < u - g < 2.7$ (see Figure \ref{fig8}). 
This color term is only used for the photometric 
calibration. Here the SDSS is used for the zeropoint only, and the SCUSS 
photometry on its native system is defined to match the SDSS at $u - g = 0$. We choose 
stars with $16.0 < u < 20.5$ to eliminate objects that are saturated or have large 
photometric errors.  Equation (\ref{equ3})  is then applied to 
transform $u$ to SCUSS $u^*$. Following Equation (\ref{equ2}), we
measure the final zeropoint $C$ by iteratively rejecting outliers. 

\begin{figure}[!h]
\centering
\plotone{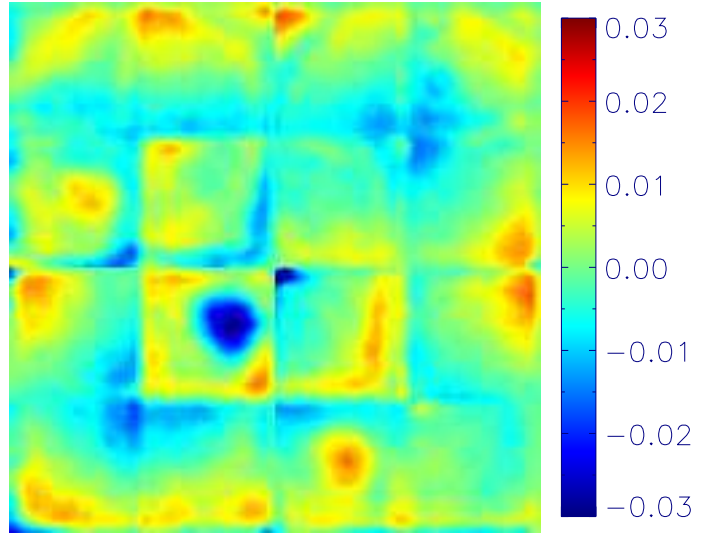}
\caption{Non-uniformity pattern used to correct the flat-fielded and 
calibrated images. The layout of these CCDs are the same as that shown in Figure \ref{fig1}. 
\label{fig9}}
\end{figure}

The CCD gain varies slightly during the observation. Furthermore, CCD gains of the  four amplifiers did not change synchronously. 
In addition, there is a photometric response non-uniformity on 
the flat-fielded images. It is possibly caused by the focal plane distortion 
and scattered light reflected in the optical system \citep{reg09,bet13}.Thus, we 
perform photometric calibration for each amplifier of each CCD. 
There are about 47 stars in each amplifier on average 
to derive the zeropoint, whose accuracy is estimated to be about 0.01 
mag. The zeropoint here is expressed in units of mag for 1ADU s$^{-1}$.
After deriving independent zeropoints for each amplifier, we find a small residual 
pattern in the sensitivity as derived from objects observed in different positions in the 
field as shown in Figure \ref{fig9}. We use this residual map to further refine the flat-field. 

\subsection{Photometric response differences of the four amplifiers}
\begin{figure*}[!htbp] \epsscale{1.0}
\subfigure{\includegraphics[width=0.5\textwidth]{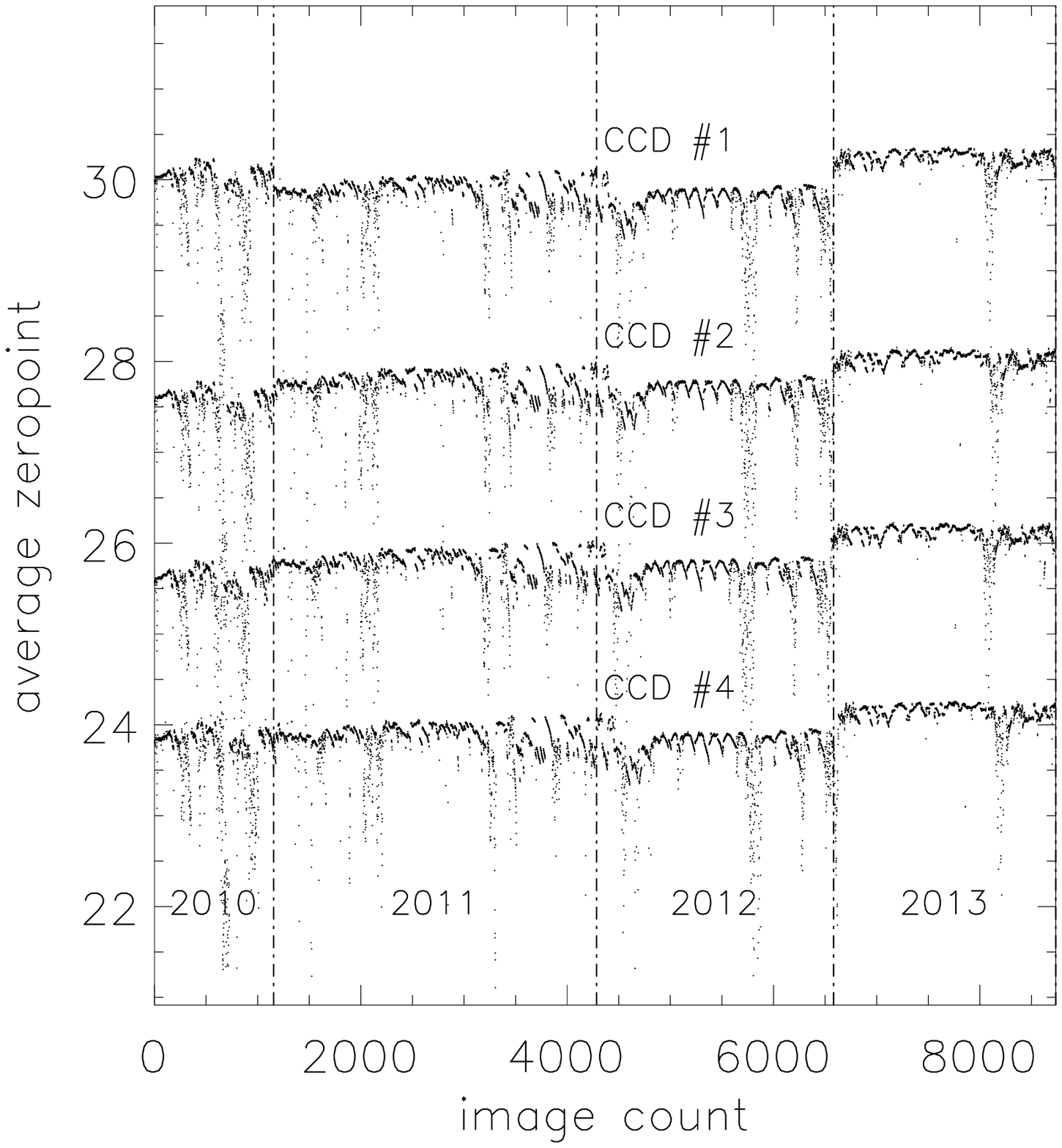}}
\subfigure{\includegraphics[width=0.5\textwidth]{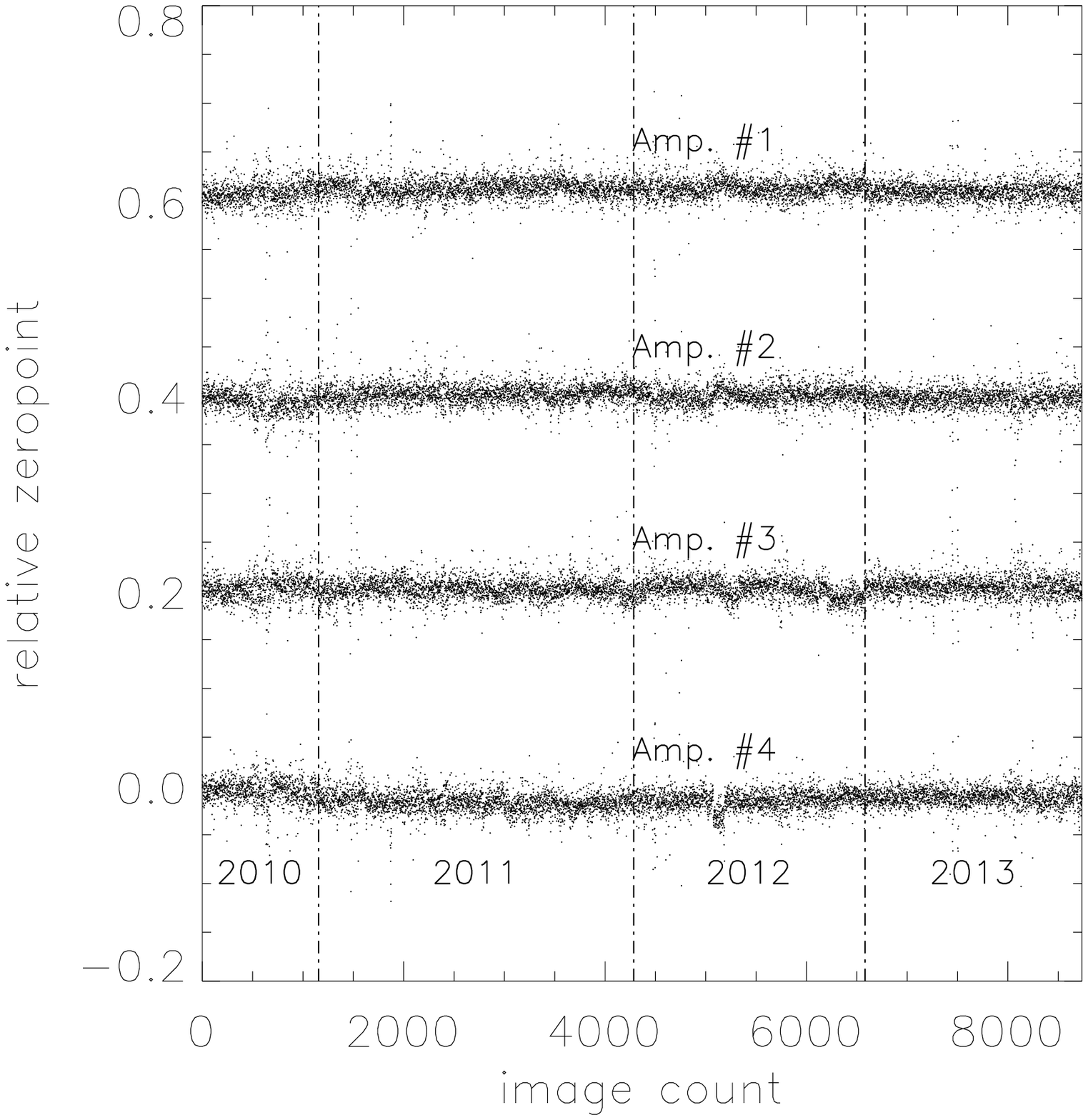}}
\caption{Left: average zeropoints of the four amplifiers for each CCD as 
a function of time. The data points correspond to the CCD images, 
which are sorted by the observation time. The zeropoints are in units 
of mag for 1 ADU s$^{-1}$. The data for CCD \#1, \#2, and \#3 are 
shifted by 2.0. The dashed lines separate the years of observations. 
Right: relative zeropoints compared with the above average for each 
amplifier of CCD \#1 as a function of time. The data for three amplifiers 
, Amp. \#1, \#2, and \#3, are shifted by 0.2.\label{fig10}}
\end{figure*}

The average zeropoints of four amplifiers are shown in the left panel of 
Figure \ref{fig10} as a function of time. Usually during the night, we scan the sky 
from east to zenith and then to west. The airmass and corresponding atmospheric extinction
first decreases and then increases. 
Therefore, zeropoints present nightly variations. From this figure, it can be 
seen that the weather conditions over 2010 are worst. The right panel of Figure \ref{fig10} 
shows the relative zeropoint variation with time for each amplifier of 
CCD \#1, compared with the average zeropoint of all amplifiers. The 
relative zeropoint variations do not keep the same value and sometimes two 
amplifiers present opposite variations. The detectors have bad qualities 
in 2010 so that the relative zeropoint changes more notably with time. 
Thus, it is critical that we obtain the 
photometric solutions independently for the four amplifiers. 

Table \ref{tab3} gives the zeropoint scatter of four amplifiers for 
each year and each CCD. The zeropoint differences among amplifiers 
originate from  gain variation, photometric response non-uniformity 
as mentioned before, and the photometric calibration error. The overall 
zeropoint scatter of four amplifiers around the averages are 
0.012$\pm$0.006, 0.016$\pm$0.007, 0.015$\pm$0.009, and 0.011$\pm$0.006 
for CCD \#1, \#2, \#3, and \#4, respectively. The general response 
differences of four amplifiers in all CCDs are less than 1.5\% 
except the CCD \#3 of 2010, which is about 2.8\%. 

\begin{table}[!h]
\footnotesize
\centering
\caption{Average zeropoint scatter of four amplifiers for each CCD \label{tab3}}
\begin{tabular}{c|cccc}
\hline
\hline
\backslashbox{year}{CCD} & \#1 & \#2 & \#3 & \#4 \\
\hline
2010&0.011$\pm$0.007&0.015$\pm$0.009&0.028$\pm$0.009&\nodata  \\
2011&0.013$\pm$0.006&0.015$\pm$0.006&0.011$\pm$0.006&0.010$\pm$0.006 \\
2012&0.013$\pm$0.006&0.013$\pm$0.006&0.010$\pm$0.006&0.011$\pm$0.006 \\
2013&0.010$\pm$0.006&0.020$\pm$0.007&0.017$\pm$0.006&0.011$\pm$0.006 \\ 
All &0.012$\pm$0.006&0.016$\pm$0.007&0.015$\pm$0.009&0.011$\pm$0.006 \\
\hline
\end{tabular}
\end{table}

\subsection{Internal calibration}
\begin{figure}[!h] \epsscale{1.0}
\plotone{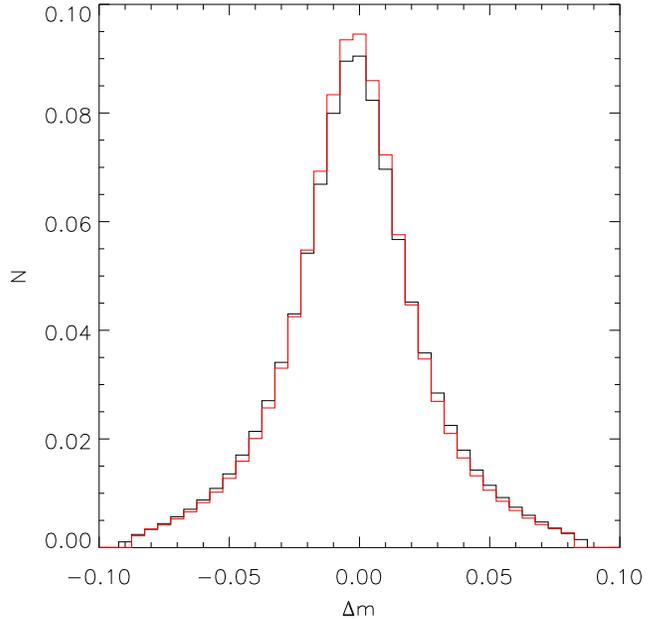}
\caption{Magnitude difference distributions between two measurements 
for objects observed more than twice. The black histogram is externally 
calibrated by the SDSS catalog, while the red one is internally 
calibrated. Only bright objects with 16 $ < u < $19 are selected.
\label{fig11}}
\end{figure}

Each SCUSS field is observed twice and overlaps with the adjacent 
fields. Thus, each CCD image can be calibrated using the common objects
cross-identified in surrounding images. The calibrating process begins with  balancing the zeropoints of images covered by the SDSS and 
then transferring the photometric solutions to the images out of the SDSS 
footprints. The calibration iterates until the whole grid of 
photometric solutions finally converges. Figure \ref{fig11} shows the 
magnitude difference distribution between measurements for objects that 
are observed twice. The histograms in black and red are the 
distributions of the same objects with the SCUSS $u$-band magnitude 
calibrated externally and internally, respectively. The dispersion by 
external calibrations is about 0.028 mag, while that by internal 
calibrations is about 0.025. It seems the internal calibration is at 
least as good as or even a little better than the external calibration.

\section{Image Quality Statistics} \label{sec6}
Both weather and instrumental status affect the image quality. The 
camera was regularly updated after each observation season, so its 
performance was improved gradually. The weather and night 
sky conditions strongly affect the imaging depth of the survey. We can measure 
the characteristics tracing the image quality, such as airmass, 
sky background brightness, seeing, and photometric zeropoints.

\subsection{Seeing}
The seeing is estimated by the FWHM measurements of isolated and 
bright point sources. The seeing distribution is presented in Figure 
\ref{fig12}a. The best seeing is 1\arcsec.2 and the overall median seeing 
is about 2\arcsec.0. The Bok telescope is located in the trough 
between two peaks of the mountain, so the wind speed is usually larger 
than other places on the same site, which has an effect on the seeing. 

\subsection{Airmass}
Figure \ref{fig12}b shows the airmass distribution of all survey 
images. The median airmass is about 1.28, and this is used to determine 
the typical $u$-band filter response as shown in Section \ref{sec-filt}. 

\subsection{Photometric zeropoint}
The variation of the photometric zeropoint mainly reflects the change 
in atmospheric transparency. The distribution of the photometric 
zeropoints is plotted in Figure \ref{fig12}c. The mean photometric 
zeropoint is 23.81 mag for 1ADU s$^{-1}$. The median zeropoints 
for airmasses of 1.0, 1.2, and 1.4 are 23.93, 23.85, and 23.75, respectively. 
There are some images with low zeropoints, most of which were observed 
when the weather was cloudy. 

\subsection{Sky brightness}
\begin{figure*}[!htbp] \epsscale{1.0} \epsscale{1.0} 
\plotone{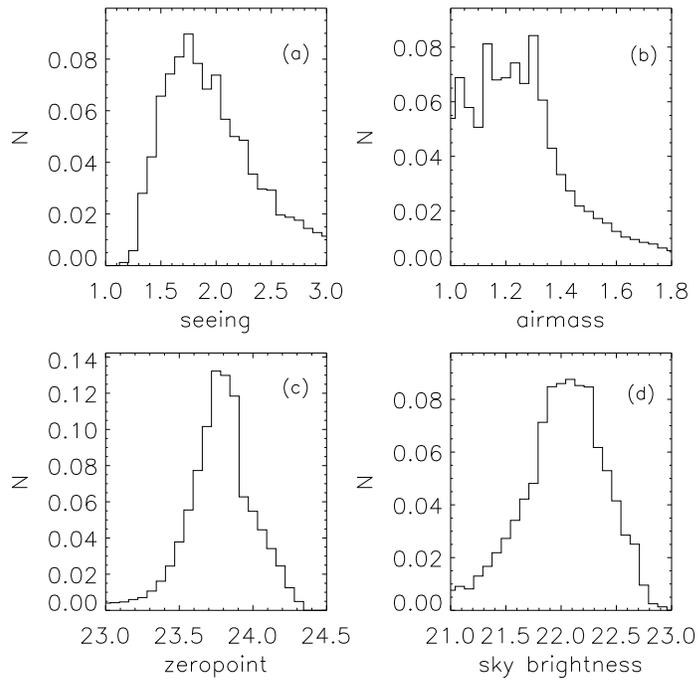} 
\caption{Histograms of seeing (in arcsec), airmass, photometric zeropoint 
	(in mag for 1 ADU s$^{-1}$), and sky background (in mag 
arcsec$^{-2}$) for all SCUSS $u$-band observations. \label{fig12}} 
\end{figure*}

The sky background brightness is an important parameter to quantify a ground-based 
observation station. Compared with other light sources, the artificial 
light pollution from nearby cities is more serious at Kitt Peak. 
The distribution of the $u$-band sky brightness at Kitt peak is 
shown in Figure \ref{fig12}d. There are some images taken when 
the moon is above the horizon. The average night sky brightness of all 
observations is about 22.05 mag arcsec$^{-2}$. The moonless median sky 
background at zenith is about 22.37 mag arcsec$^{-2}$, which is comparable 
to that of the Apache Point site (22.1 mag arcsec$^{-2}$). Note that the 
calibrated sky brightness uses the frame zeropoints that implicitly include 
atmospheric extinction, so the actual sky brightness is darker because 
the $u$-band atmospheric extinction coefficient is about 0.5, 
as estimated with the observations taken during photometric nights. 

\section{Image Stacking} \label{sec7}
\subsection{Quality Control}
Most SCUSS fields are observed under good weather and moderate 
seeing conditions. There are some cases with bad image quality: 
(1) high sky background due to observations during  astronomical 
twilight or when moon was up; (2) large seeing due to strong wind; (3) low 
atmospheric transparency due to cirrus in the FOV; 
(4) bad focus of the CCD camera. In 2013, we spent one and half observation 
runs re-observing most bad-quality fields. 

To ensure the homogeneity of the imaging depth, and completeness of the 
SCUSS survey, we keep only the images with seeing $<$3\arcsec.0, sky 
background ADU $<$500, and photometric zeropoint $>$22.56 mag. About 92.6\% 
of the survey area is covered by these images. For the remaining area, we take 
images with seeing $<$3\arcsec.0 (3.6\% area), regardless of the sky brightness and 
photometric zeropoint.  If none are available, we take all remaining images,  which 
covers about 3.8\% of the total area. 

\subsection{Image Resampling and Stacking}
Based on the central coordinates of each field, we stack related 
single-epoch images to form a combined image. We first project and resample 
the single-epoch images to a grid with a fixed pixel scale of 0\arcsec.454. 
The grid has 8640$\times$8200 pixels, covering a sky area of 
1{\arcdeg}.090$\times1{\arcdeg}.034$. If a single-epoch images contributes less than 
less than 128$\times$128, it is not included 
in the stacking process. For each pixel of the grid, there are more than 
four related pixels in each single-epoch images. These pixels are regarded 
as having the same size after being flat-fielded. We calculate the fractions 
of these pixel areas that are covered by the grid pixel and sum them to 
conserve flux.  

\begin{figure*}[!htbp] \epsscale{1.0}
\subfigure{\includegraphics[width=0.5\textwidth]{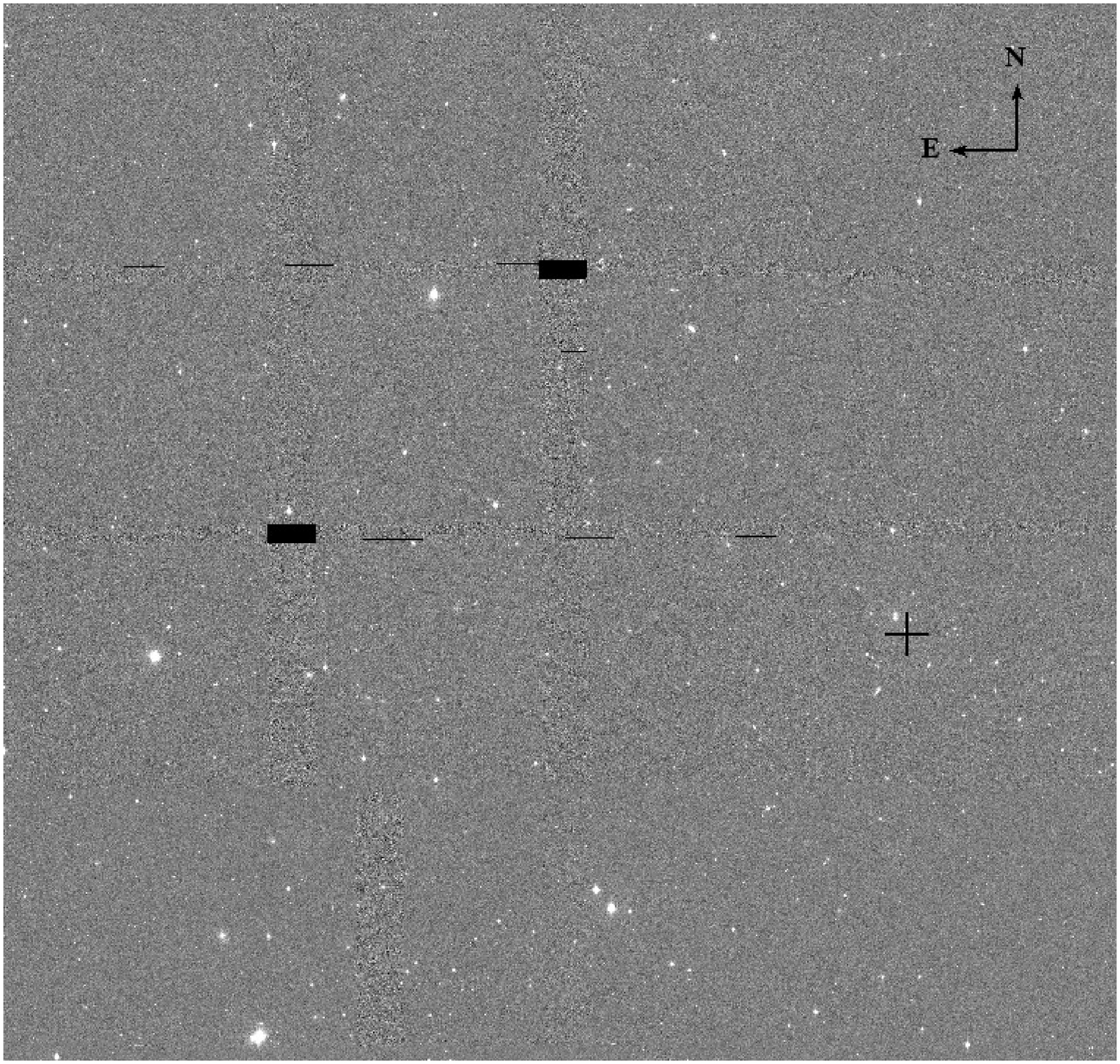}}
\subfigure{\includegraphics[width=0.5\textwidth]{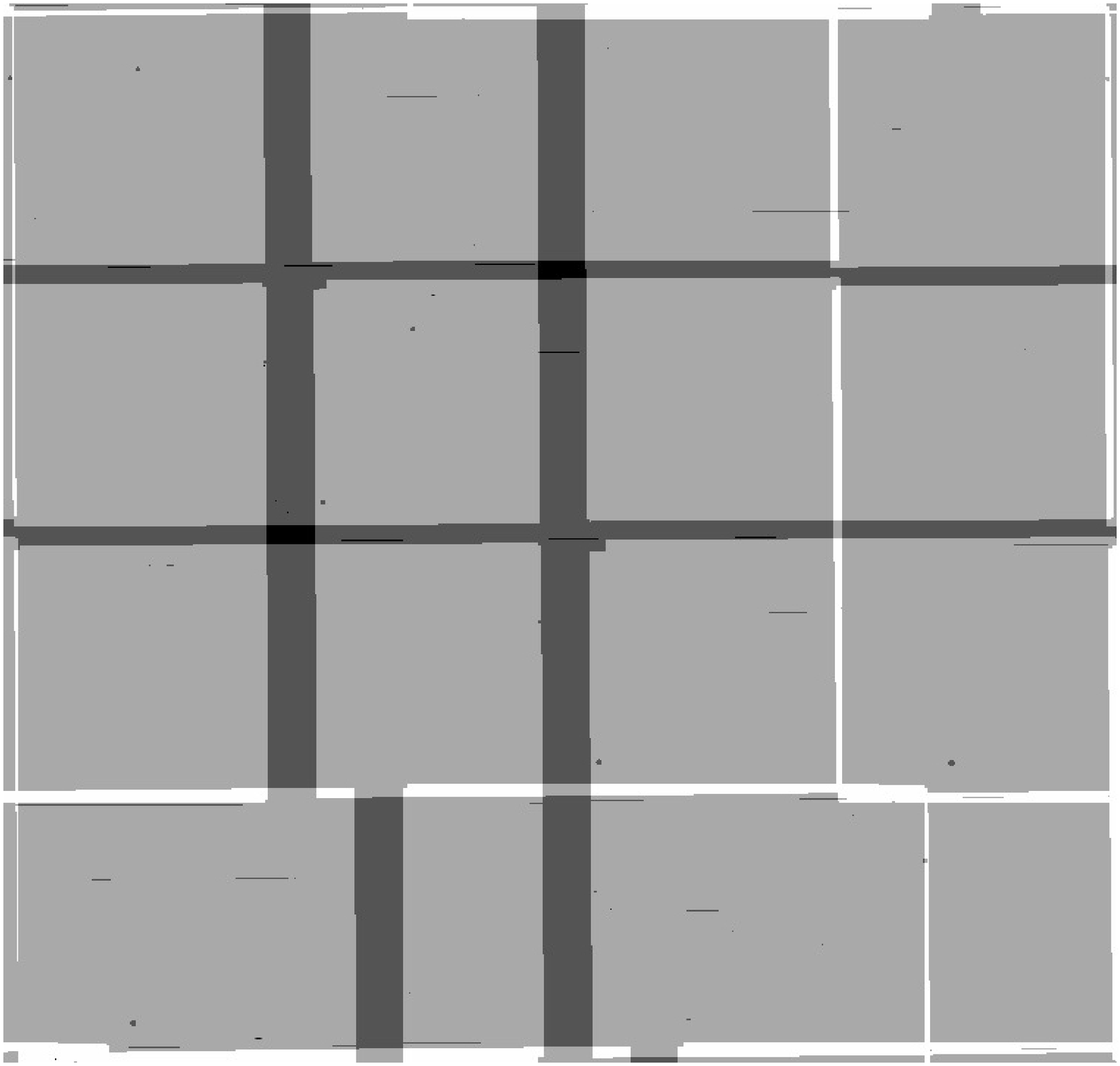}}
\caption{Examples of a stacked image (left) and its mask image (right). 
The central coordinate is $\alpha = 44\arcdeg.7536$, $\delta = 6\arcdeg.1074$.
North is up and east is left. A galactic cluster of Abell 400 (marked 
as a plus symbol in the left plot) happens to appear in this area. The 
mask image shows the exposure numbers. There are two black boxes 
presenting no observations. Most of the area has two exposures, a small 
part shows more than two exposures, and some CCD-gap areas have only 
one exposure. The small horizontal lines are caused by the missing CCD rows 
that were not recorded by the camera controller during 2011 and 2012. 
\label{fig13}}
\end{figure*}

We subtract the sky backgrounds from the resampled images. Their 
photometric zeropoints are converted to linear flux weights. The remaining 
signal after removing backgrounds are weighted by these weights and then 
co-added. If there are more than three pixels involved in stacking, 
a cosmic-ray rejection is implemented using a sigma clipping algorithm. 
We redo the flux calibration for each stacked image with the SDSS DR9 
catalog to derive the final photometric zeropoint. This zeropoint is
approximately 29 mag for 1 DN. In addition, a mask image for each field 
is also generated. Each mask pixel presents the number of images that 
are actually involved in the flux co-adding. The mask value is reduced by 
one for each epoch in which a pixel is bad, saturated, or blank. 
Figure \ref{fig13} shows a typical stacked image and its mask image. Most of 
the stacked image has two exposures, an overlapping area has more 
than three exposures, and there are some small holes located in the CCD 
gaps that are blank. Some of the CCD gaps have only one exposure, so
the depth is about 0.75 mag shallower in those locations.  

\section{Photometry} \label{sec8}
The SCUSS photometric pipeline generates comprehensive catalogs 
using multiple photometric techniques so that different users can choose the one that
suits their needs. Aperture, automatic aperture, point-spread function (PSF), and model
photometry are applied to both SCUSS stacked and single-epoch images. The 
model photometry is consistent with the SDSS model photometry that utilizes
the SDSS $r$-band model shape parameters to measure brightnesses on the SCUSS $u$-band 
images.
\subsection{Source detection}
Source detections for astronomical science images are never complete, 
especially at the fainter magnitude end. Morphologies of extended 
sources in $u$ band are more fragmented and diffuse than those in other 
optical bands. Therefore, many  fainter sources with low surface 
brightness might be missing. In addition, most of the science projects 
based on SCUSS data also need to use  deeper and redder other bands 
from other large-scale surveys. More than 3/4 of the SCUSS area is 
covered by the SDSS. Our sources 
include both SDSS detected objects and detections unique to our SCUSS 
images. For the area not covered by the SDSS, we perform photometry for 
only SCUSS detections. The resulting catalog within these areas will be useful 
to match with other wide imaging surveys (e.g. Pan-STARRS).
 
The SDSS objects with any one of the PSF, Petrosian, model, and 
CModel $ugriz$-band magnitudes brighter than 23.5 mag are selected. 
SExtractor is used to detect objects in SCUSS stacked images, and these  
are matched with the SDSS objects.  The mismatched objects are 
rematched with SDSS catalogs with magnitudes fainter than 23.5 mag in order 
to find missing SDSS objects. The rest of the objects are SCUSS unique 
detections. 

\subsection{Aperture photometry by DAOPHOT}
Circular aperture photometry is a simple procedure to measure 
magnitudes of sources. The core code of the aperture photometry in 
DAOPHOT is utilized to measure aperture magnitudes \citep{ste87}.
We use 12 apertures with radii ranging from 1{\arcsec}.4 to 18{\arcsec} (from 3 to 
40 pixels; see Table \ref{tab4}).

\begin{table}[!h]
\small
\centering
\caption{Aperture radii for the aperture photometry \label{tab4}}
\begin{tabular}{ccc}
\hline
\hline
No. & Radius in Pixels & Radius in Arcsec \\
1  & 3  & 1.36 \\
2  & 4  & 1.82 \\
3  & 5  & 2.27 \\
4  & 6  & 2.72 \\
5  & 8  & 3.63 \\
6  & 10 & 4.54 \\
7  & 13 & 5.90 \\
8  & 16 & 7.26 \\
9  & 20 & 9.08 \\
10 & 25 & 11.35 \\
11 & 30 & 13.62 \\
12 & 40 & 18.17 \\
\hline
\end{tabular}
\end{table}

\begin{figure*}[!htbp] \epsscale{1.0} 
\plotone{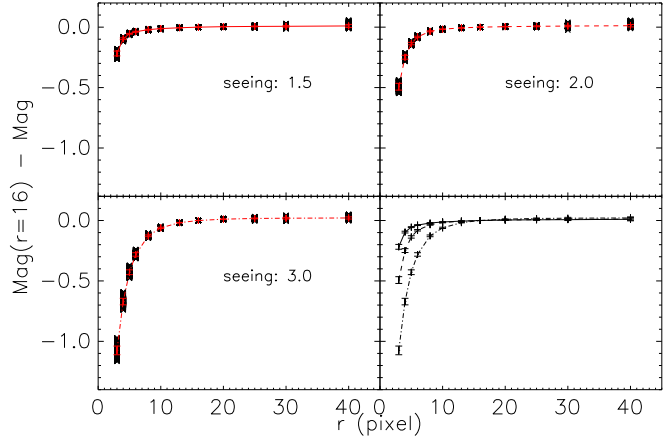} 
\caption{Growth curves for aperture corrections at different seeing.
The black pluses are growth curves for about one hundred isolated and
point-like objects. The red lines with error bars are the medians and 
standard deviations of the corrections for different apertures. The 
last panel shows the median growth curves in the previous three 
panels in order to present the impact of seeing. \label{fig14}} 
\end{figure*}

It is important to apply aperture corrections to aperture magnitudes 
due to flux loss within a finite aperture size. For smaller apertures and larger seeing, this is
especially important. Figure \ref{fig14} presents growth curves under different 
seeing conditions. The growth curve is calculated as the magnitude 
difference between one of 12 apertures and the 7\arcsec.3 radius aperture
as function of aperture radius. This reference aperture is the 
one used for photometric flux calibration. We choose isolated 
and point-like objects with photometric errors in all apertures less than 
0.05 mag to derive the growth curve. Outlier objects are eliminated when the median growth curve 
is calculated. 

\subsection{Automatic photometry by SExtractor}
\begin{figure*}[!htbp] \epsscale{1.0} 
\plotone{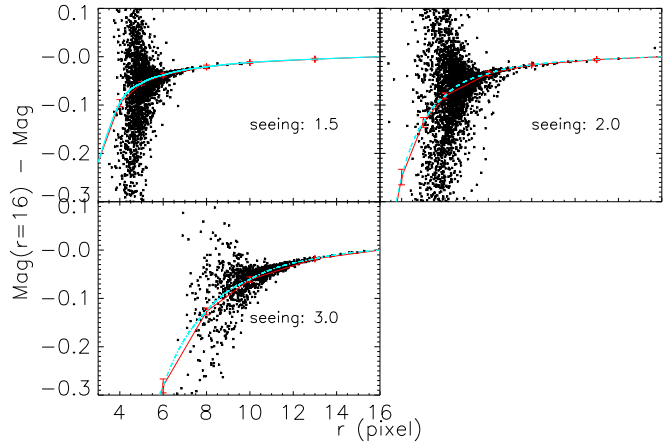} 
\caption{Difference between the automatic and circular aperture magnitudes 
of 7{\arcsec}.3 radius as a function of the characteristic radius. Black dots 
are isolated point sources. Growth curves for corresponding circular aperture 
corrections shown in Figure \ref{fig14} are overlaid in red. The cyan 
curves are the cubic spline interpolation of those growth curves. 
\label{fig15}}
\end{figure*}

SExtractor \citep{ber96} provides precise magnitudes 
of sources using automatic aperture photometry, generating so-called "automatic 
magnitude", which is motivated by the Kron algorithm \citep{kro80} 
\footnote{https://www.astromatic.net/software/sextractor}. 
An elliptical aperture is automatically determined for each object to 
integrate the flux. Although most of the flux is 
expected to lie within the elliptical aperture and the flux loss should be almost 
independent of the source magnitude, we discover that in fact,  the flux loss does change with 
both the source magnitude and seeing. The magnitudes of objects that 
are brighter or objects observed with worse seeing have more corrections.  
We consider an equivalent circle whose area is equal to that of 
the ellipse. Thus, the radius of this circular aperture is equal to 
the root of the product of the semi-major and semi-minor axis lengths. 
We call this circular radius the characteristic radius, which 
has the same meaning as the aperture size in normal aperture photometry discussed above.
Figure \ref{fig15} presents the aperture corrections for the 
automatic magnitudes with different seeing. The black dots are point-like 
and isolated objects detected by SExtractor. They lie along with the growth 
curve of the aperture correction described in the previous section. Thus, 
an interpolation from the curve is good enough for correcting automatic 
magnitudes.

We also perform Petrosian-like photometry on SCUSS images by SExtractor. 
The Petrosian aperture is very similar to the Kron one. They share 
the same position angle and ellipticity. The Petrosian aperture radius 
is determined by the ratio of the isophotal brightness at a certain radius and 
average surface brightness within this radius.
The ratio is set to be 0.2 and  the corresponding  Petrosian radius is larger than the Kron radius. 
The aperture corrections for Petrosian magnitudes can be 
estimated by the same way as mentioned above.

\subsection{PSF photometry}
The PSF is obtained using PSFEx\footnote{http://www.astromatic.net/software/psfex}
\citep{ber11}. The form of the PSF in PSFEx is expressed as a linear 
combination of basis vectors. The pixel basis is selected in our PSF 
modeling. The spatial PSF variation on the focal plane usually shows a 
smooth profile, which can be modeled by a low-order polynomial. For 
SCUSS single exposures, a second-order polynomial is good enough to describe 
the PSF variation over the CCD plane. For SCUSS stacked images, a 
seventh-order polynomial is used to describe the complexity of the image quality.

\begin{figure*}[!htbp]\epsscale{1.0}
\plotone{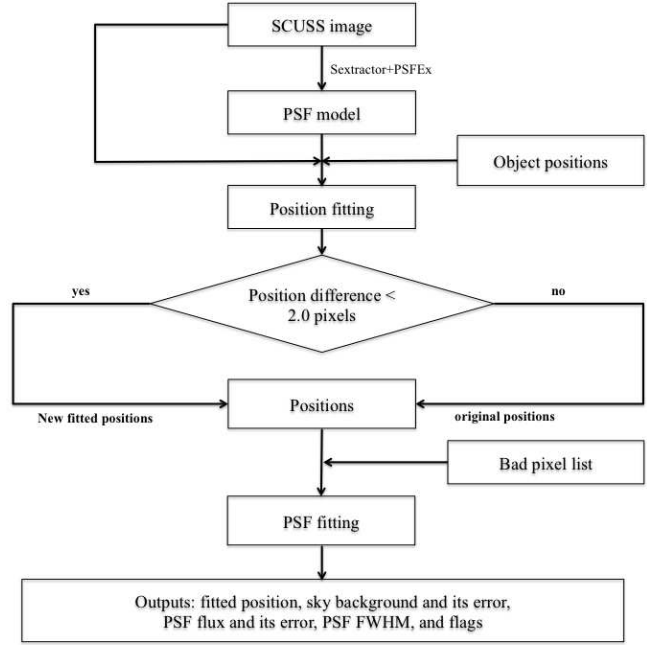} 
\caption{Flow chart of our PSF photometry pipeline. Firstly, the PSF model is extracted 
by combining SExtractor and PSFEx. Secondly, the coordinate for each object is 
fitted by a Gaussian function. Thirdly, the image of each object with the sky 
background subtracted is fitted by the local PSF profile. Finally, the 
photometric results together with the fitted positions and flags are given as outputs.  
\label{fig16}}
\end{figure*}

A code is specially designed to measure the PSF magnitudes at known object 
positions. The code takes the SCUSS image, SDSS and SCUSS-only object 
positions, the bad-pixel list, and the PSF model derived by PSFEx as inputs 
and then outputs the Gaussian-fitted position, local sky background and its 
error, local PSF FWHM, fitted PSF integrated flux and its estimated error, 
and a flag tagging the status of each object. Figure \ref{fig16} shows 
the flowchart of the PSF photometry code. The local sky background for each 
object is measured using the pixels at $r > 7.5\times$FWHM. Here, the FWHM 
is the full width at half maximum of the local PSF model. Outliers, such 
as cosmic rays and signals from real objects, are iteratively rejected by 
a sigma-clipped algorithm. After the sky background is subtracted, the 
object position on the CCD is fitted by a two-dimensional Gaussian function. 
The position is allowed to shift because the coordinates in SDSS redder bands are slightly different 
from those in the SCUSS $u$ band due to atmospheric refraction and 
star proper motions. The pixels centered at the initial position with $r < 2.5\times$FWHM are 
considered in the calculation. If the new fitted position is more than 2.0 
pixels away from the old one, the following PSF photometry will be performed 
at the original position. The PSF model is interpolated to the same pixel 
scale of the CCD image and fitted to the fluxes of the object pixels within 
$r = 2.0 \times$ FWHM. Flags are also provided by our PSF photometry to show 
the reliability. They are coded in decimal and expressed as a sum of powers 
of 2: (1) CCD artifacts; (2) bad pixels; (4) including saturated pixels; (8) 
contaminated by neighbors; (16) near image edges.

\begin{figure*}[!htbp] \epsscale{1.0}
\plotone{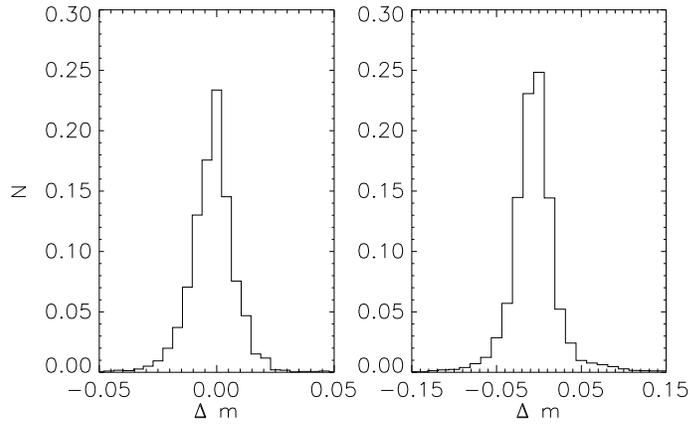}
\caption{Comparisons of our PSF photometry with DAOPHOT. Left is 
the comparison for common bright objects with $16 < u < 18$ and right is 
the one for common faint objects with $21< u < 22$.  
\label{fig17}}
\end{figure*}

We compare our PSF measurements with one of the popular PSF photometry software packages  
(DAOPHOT), which is a widely used package for accurate stellar photometry designed 
to deal with crowded fields. We use DAOPHOT to find objects in one single-epoch 
image and at the same time give the PSF magnitude measurements. Then, our code 
performs PSF photometry for those objects with the PSF model derived by PSFEx. 
The magnitude comparison is shown in Figure \ref{fig17}. Two groups of objects 
are chosen: a brighter one with $16 < u < 18$ mag (left in Figure \ref{fig17}) and a 
fainter one with $21 < u < 22$ mag (right in Figure \ref{fig17}). The scatter
of the PSF magnitude differences between our code and DAOPHOT are 0.009 mag for 
the bright group and 0.021 for the faint group, respectively. Our PSF photometry 
is consistent with DAOPHOT, although the PSFs used by these two 
techniques are derived in different ways.  

\subsection{Model photometry}
The SDSS also provides a type of model magnitude, or "modelMag," which can 
measure the unbiased colors of galaxies in the absence of color gradients
through equivalent apertures in different bands. The "modelMag" is 
generated by choosing a shape model, either a deVaucouleurs or exponential profile, 
based on its best-fit likelihood in the SDSS $r$ band,  then convolving them with 
seeing in other bands, and finally forcing magnitude measurements with the same aperture shape.   

We divided SDSS objects into two groups, point sources and extended sources, 
based on the SDSS classification. For extended sources, we 
construct their theoretical 2D models with the effective radii, axis 
ratios, and position angles from the SDSS measurements. These models are convolved with local 
PSF profile derived by PSFEx. For point sources, we use local PSF from 
PSFEx directly as their models. In addition, the extended sources with 
small sizes or low brightnesses (effective radii less than 0.5 arcsec or 
SDSS $r$-band magnitude fainter than 23.5 mag) are also treated as 
point sources.

After the models are constructed, model amplitudes are calculated as the ratios 
of the models and raw SCUSS fluxes. The pixels within $r = 1.0 \times$ FWHM are 
considered in the calculation, which is optimized for faint SDSS extended 
sources. According to the amplitudes, we then compute the corresponding 
deVaucouleurs and exponential magnitudes. Therefore,  the "modelMag" in 
SCUSS $u$ band can be estimated by the SDSS $r$-band deVaucouleurs and 
exponential profiles. The SCUSS model magnitude is aperture-corrected to 
make the magnitudes of bright point sources ($16 < u < 20$) equal to the SDSS $u$-band model 
magnitudes.  

\subsection{Co-added photometry} \label{sec-6.6}
The stacked images are composed of single-epoch images taken with 
different observational conditions. The seeing varies between each single-epoch image, 
which makes the PSF profile of the stacked image 
quite complicated. Subsequently, the fraction of flux loss due to finite 
apertures for objects in different parts of the stack images might be 
different. In addition, the PSF profile cannot be perfectly determined 
unless the seeing of related single-epoch images is similar enough. 
On the other hand, the PSF profile of a single image varies smoothly 
and it is much easier to model. We can first perform photometry for an 
object observed with different exposures and then co-add its fluxes 
to generate different co-added magnitudes at the catalog level.

The aperture photometry for single-epoch images is obtained by DAOPHOT 
with 12 apertures as defined before. Aperture corrections are applied 
to the resulting aperture magnitudes. Corresponding corrected fluxes 
are weighted by the errors and averaged to generate the co-added 
aperture magnitudes. 

The PSF photometry for single-epoch images is performed using our PSF 
fitting code. The PSF magnitudes are also aperture-corrected by 
comparing with the aperture magnitudes in a 7\arcsec.3 radius. The 
co-added PSF magnitudes are calculated by averaging the fluxes weighted 
by their errors. 

We perform the model photometry for single-epoch images and measure 
the exponential and deVaucouleurs magnitudes. We adopted model magnitude with  
a higher SDSS $r$-band likelihood. Aperture corrections are applied to make model 
magnitudes equal to the SDSS model magnitudes in the case of unresolved 
objects. The co-added model magnitude is calculated by averaging the 
model fluxes weighted by their errors.

\section{Some photometric comparisons} \label{sec9}
\begin{figure*}[!htbp] \epsscale{1.0}
\centering
\plotone{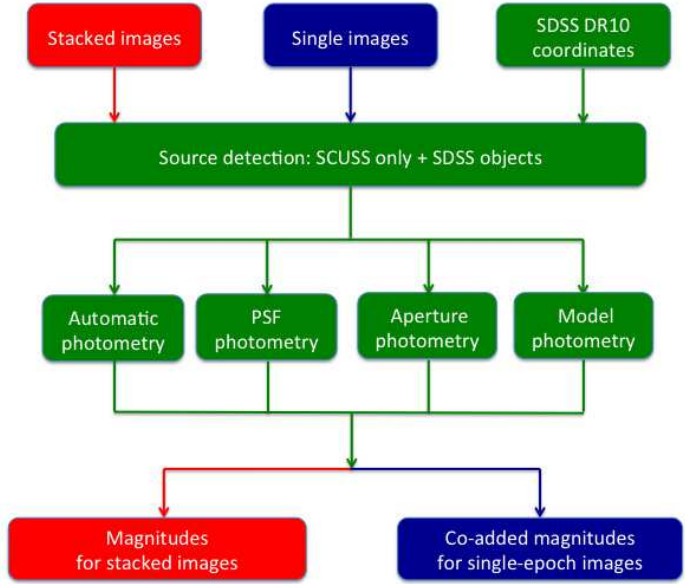}
\caption{SCUSS photometric scheme. The output magnitudes include the 
automatic, Petrosian, aperture, and PSF magnitudes for the stacked 
images and co-added aperture and PSF magnitudes for the single-epoch 
images. \label{fig18}}
\end{figure*}

The photometric methods described above are applied to stacked and 
single-epoch images. Figure \ref{fig18} illustrates the general photometric 
scheme. The  detections are based on both SDSS and SCUSS images. We perform  
aperture, PSF, model, and automatic photometry for the stacked images. 
Note that the automatic magnitudes are only based on objects solely detected 
by SExtractor. We also obtain co-added aperture, PSF, and model magnitudes from 
photometry of  
single-epoch images. Since point sources can be best modeled while extended 
sources are much more complicated, the photometry of point-like objects is better suited for comparisons of our different photometric methods. 
Thus, the comparisons of point sources are presented for most cases below and, for comparison 
with SDSS, we correct the magnitude with the SCUSS/SDSS color term.

\subsection{Photometry of stacked images}
We compare the different SCUSS photometric measurements for stacked images 
with the SDSS PSF or model magnitudes of point sources in Figure \ref{fig19}. 
The SCUSS model magnitude is compared with the SDSS $u$-band model 
magnitude, while other magnitudes are compared with the SDSS PSF magnitude. 
We choose the 5 pixel (2\arcsec.27) to present the aperture magnitude in this figure. For the smallest apertures, 
the aperture magnitudes might be problematic due to inhomogeneous image quality 
in some stacked images. Table \ref{tab5} gives the magnitude offset and scatter in two 
magnitude intervals: $16 < u < 19$ mag and $20.5 < u < 21.5$ mag (i.e. $\sim$21 mag). 
The automatic magnitude is best among 
all photometric magnitudes for stacked images, with a scatter of 0.033 for bright 
stars and 0.178 for faint stars at $u\sim21.0$ mag. 

\subsection{Co-added photometry for single images}

\begin{figure*}[!htbp] \epsscale{1.0}
\plotone{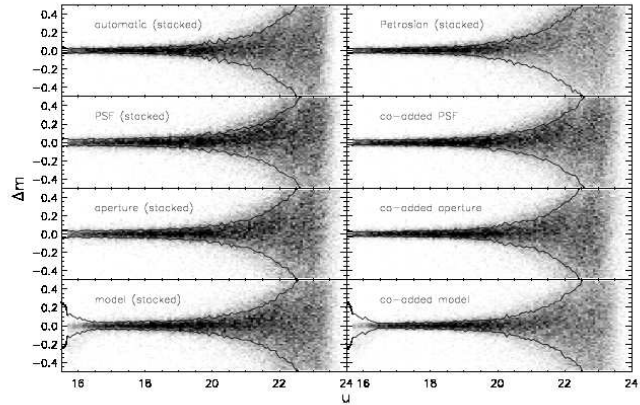}
\caption{Magnitude differences between different SCUSS magnitudes and the 
SDSS PSF or model magnitudes of points sources as functions of SCUSS 
magnitudes. The first-row panels and the rest of the left panels are comparisons of 
automatic, Petrosian, PSF, and aperture magnitudes for the stacked images. 
The right panels from the second row to bottom are comparisons of co-added 
PSF, aperture, and model magnitudes for the single-epoch images. The 
aperture magnitude shown here is the one with an aperture radius of 
4{\arcsec}.5. The two lines above and below zero show $\pm1\sigma$ scatter
around the average. 
\label{fig19}}
\end{figure*}

The comparisons of the co-added PSF, aperture, and model magnitudes 
with the SDSS PSF or model magnitudes are shown in the right panels of 
Figure \ref{fig19} and the last three rows of Table \ref{tab5}. The 
co-added PSF magnitude performs the best for point sources among all the magnitude 
types. The scatter is 0.033 
for brighter sources and 0.174 for fainter sources at $u \sim 21.0$ mag. 
The co-added aperture magnitude is also adequate if a proper aperture radius 
is considered. The aperture size should be chosen according to the object 
type, object brightness, signal-to-noise requirement, etc. 

\begin{table}[!h]
\centering
\footnotesize
\caption{Magnitude offset and scatter of point sources for the SCUSS magnitudes compared with the SDSS PSF or model magnitudes\label{tab5}}
\begin{tabular}{l|cc|cc}
\hline
\hline
\multirow{2}{*}{Magnitude} & \multicolumn{2}{c|}{$16< u <18$} & \multicolumn{2}{c}{$20.5< u <21.5$} \\ \cline{2-5} 
                  &  Offset & Scatter & Offset & Scatter \\
\hline
Automatic (stacked)         &   0.001 & 0.033 &  0.008 & 0.178\\ 
Petrosian (stacked)         &   0.002 & 0.033 &  0.065 & 0.200\\
PSF (stacked)               &  -0.003 & 0.047 &  0.025 & 0.175\\
Aperture (stacked)          &  -0.002 & 0.044 &  0.017 & 0.176\\
Model (stacked)             &  -0.006 & 0.059 &  0.004 & 0.177\\
Co-added PSF       &  -0.004 & 0.033 & 0.021 & 0.174\\
Co-added aperture  &   0.003 & 0.034 & 0.010 & 0.177\\
Co-added model     &  -0.003 & 0.049 & 0.003 & 0.175\\
\hline
\end{tabular}
\end{table}
\subsection{Comparisons with the CFHTLS deep $u$ band}
The Canada-France-Hawaii Telescope Legacy Survey \citep[CFHTLS;][]{ast06} used the wide-field 
optical imaging camera  on CFHT to obtain deep multicolor photometry over wide areas over 
5 years.The photometric system is similar to that of the SDSS, except the $u$ 
filter. As one component of the CFHTLS, the wide synoptic survey covers 155 
square degrees in four patches down to $i=$24.5. The $u$-band magnitude limit 
in the wide survey reaches 25.2 mag, which is much deeper than the SCUSS $u$ 
band. Thus, the catalogs of the wide survey can be regarded as a reference 
when comparing the SCUSS photometry with that of the SDSS. There are 
two CFHTLS wide regions, W1 and W4, overlapping with the SCUSS footprints. 
The W4 field ($\alpha = \mathrm{22^h13^m18^s}, \delta = \mathrm{+01^d19^m00^s}$)
has an area of 25 deg$^2$ fully covered by the SCUSS and it is located 
at a higher declination, where the SCUSS data quality is more typical than 
it is in the W1 field. We make some photometric comparisons by using the W4 
catalogs.

\begin{figure*}[!htbp] \epsscale{1.0}
\plotone{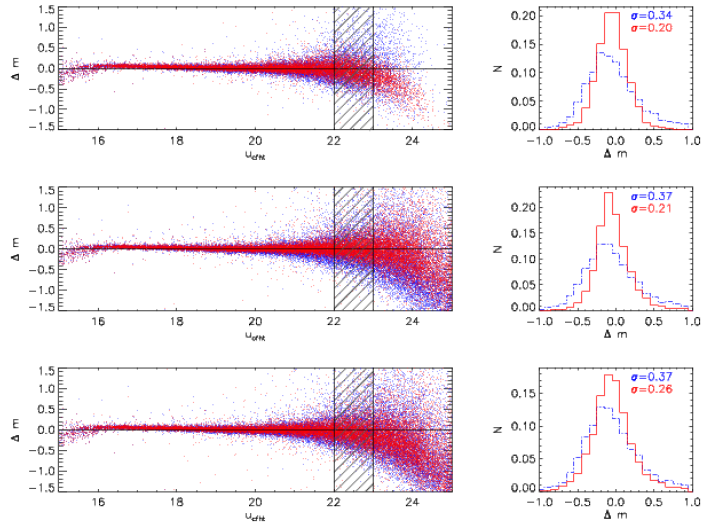}
\caption{Photometric comparisons of point sources between the SCUSS (in red) 
and SDSS (in blue) with the CFHTLS W4 data as a reference. The SDSS magnitude 
is the PSF magnitude and the CFHTLS magnitude is the automatic magnitude. The 
left panels are the magnitude differences between the SCUSS or SDSS and the 
CFHTLS as a function of the CFHTLS $u$-band magnitude, and the right ones are 
the normalized distributions of the differences at the faint magnitude end 
(the hatch area: $22 < u < 23$ mag). From top to bottom, they are comparisons 
of three types of SCUSS magnitudes: the automatic, co-added PSF, and co-added 
aperture magnitudes, respectively. Only about 20,000 objects are randomly 
selected in this figure in order to avoid crowding. The color texts in 
the right panels gives the magnitude scatter of the SCUSS and SDSS, which 
corresponds to the 68.3\% confidence level. 
\label{fig20}}
\end{figure*}

Figure \ref{fig20} shows the photometric comparisons of point sources between 
the SCUSS and SDSS with the CFHTLS wide data as a reference. The left panels 
in this figure present the magnitude difference between the SDSS (in blue 
points) or SCUSS $u$-band magnitude (in red points) and the CFHTLS automatic 
magnitude as a function of the CFHTLS $u$ magnitude. The SDSS magnitude is 
the PSF magnitude and the SCUSS magnitudes are automatic, co-added PSF, and 
co-added aperture (2\arcsec.27) magnitudes from top to bottom, respectively. 
These three types of SCUSS magnitudes are considered to be the best flux measurements 
for point sources. The number of objects for the automatic magnitude is less than those 
of the other two magnitude types (mainly at the faint end) due to different  
source detection by SExtractor itself. Both the SCUSS and SDSS $u$-band 
magnitudes are converted to the CFHTLS photometric system by the color term 
as indicated in the CFHTLS webpage \footnote{http://cfht.hawaii.edu/Instruments/Imaging/MegaPrime/generalinformation.html}: $u_\mathrm{CFHT} = u_\mathrm{SDSS/SCUSS} - 0.241(u_\mathrm{SDSS/SCUSS} - g_\mathrm{SDSS})$. The SDSS 
scatter for brighter sources is similar to the SCUSS scatter, while it 
is much larger than the SCUSS scatter at the faint magnitude ends. The 
histograms in the right panels of Figure \ref{fig20} show the magnitude 
difference distributions with $22< u_\mathrm{CFHT} < 23$. The scatter for the SCUSS 
and SDSS are also presented in these panels. The photometric accuracy of 
the SCUSS is much better than that of the SDSS for fainter sources due to 
deeper imaging.

\subsection{Photometry for extended sources}
Extended sources are much more complicated, especially in ultraviolet 
bands, since their morphologies look much more fragmented than in redder 
bands. By comparing different magnitude measurements on stacked images 
with the SDSS model magnitude, we calculate a scatter at $u \sim 21$ of about  
0.2, 0.31, 0.4, 0.23 for automatic, aperture (2\arcsec.27), PSF, 
and model magnitudes, respectively. The co-added photometric magnitudes 
give similar results. The automatic magnitudes seems to be the best for extended 
sources, but they only describe the brightnesses of SCUSS-detected objects. 
The model and aperture magnitudes are also adequate. The PSF photometry fails 
to measure the magnitudes of galaxies.

\subsection{Guidelines to use magnitudes}
The choice of photometric technique is dependent on the science one wants to do. 
Here we present some general guidelines. More than 90\% of the stacked images are assembled from single-epoch 
images with consistent image quality. The photometry on these images is as 
good as the corresponding co-added photometry. However, as the background noise 
in single-epoch images is larger than that in stacked images, there are about 23\% more 
objects with available magnitude measurements in stacked images than with available co-added magnitudes.

The automatic magnitude performs as well as the PSF and model magnitudes because it can 
adaptively fit elliptical apertures to both point and extended sources with similar flux loss. 
It can be regarded as a universal magnitude for both point-like 
and extended sources. But unlike other photometric methods, the automatic 
photometry is based on the objects detected on SCUSS images, which are about 35\% of 
SDSS objects with available SCUSS $u$-band fixed-parameter measurements. For point 
sources like quasars and stars, the PSF and aperture magnitudes with 
appropriate aperture sizes are recommended. The aperture size needs 
to be determined based on scientific objectives. For nearby galaxies with extended morphological 
structures, the automatic magnitude and aperture magnitude are good choices. 
The SCUSS model magnitude is defined the same as the SDSS "modelMag". When combining 
magnitudes of other SDSS bands to measure the colors of extended sources, 
it is better to use the model magnitude.

\section{Summary} \label{sec10}
The SCUSS survey is a wide-field $u$-band sky survey in the Southern Galactic cap. 
The survey used the Bok telescope on Kitt Peak and the filter is close to the SDSS 
$u$ band. The survey observations were completed by the end of 2013 and the total 
area is about 5000 deg$^2$. This paper describes the detailed data reduction 
dedicated to the survey, including basic image processing which has some special features
related to the detectors, astrometric and photometric calibrations, and 
photometry. The general astrometric error is about 0\arcsec.13. The SCUSS photometric 
calibration is tied to the SDSS catalogs and is performed for each amplifier 
of a CCD due to gain and weather variations. 

We apply different photometric techniques to the stacked images, including automatic 
photometry by SExtractor, aperture photometry by DAOPHOT, PSF photometry, and model 
photometry. Our PSF photometry with more controllable parameters is   
consistent with DAOPHOT. The model photometry is similar to the SDSS "ModelMag", which 
uses the SDSS $r$-band model-derived shape parameters and SCUSS PSF profiles to make consistent 
and unbiased model magnitude measurements. We perform photometry on stacked images and also 
on single-epoch images, from which co-added photometry is derived.
More than 90\% of stacked images are assembled from single-epoch 
images with consistent quality.  There are about 23\% more objects with available magnitudes on stacked images 
than with available co-added magnitudes. The photometry on these 
stacked images is as good as the co-added photometry. However, for the rest of the stacked images, 
their photometry is worse than the co-added photometry due to uneven image quality.

\acknowledgments
We thank the referee for his/her thoughtful comments and insightful suggestions that greatly improved our paper. This work is supported by the National Natural Science Foundation of China (NSFC, Nos. 11203031, 11433005, 11073032, 11373035, 11203034, 11303038, 11303043) and by the National Basic Research Program of China (973 Program, Nos. 2014CB845704, 2013CB834902, and 2014CB845702). Z.Y.W. was supported by the Chinese National Natural Science Foundation grant No. 11373033. This work was also supported by the joint fund of  Astronomy of the National Nature Science Foundation of China and the Chinese Academy of Science, under Grant U1231113.

The SCUSS is funded by the Main Direction Program of Knowledge Innovation of Chinese Academy of Sciences (No. KJCX2-EW-T06). It is also an international cooperative project between National Astronomical Observatories, Chinese Academy of Sciences, and Steward Observatory, University of Arizona, USA. Technical support and observational assistance from the Bok telescope are provided by Steward Observatory. The project is managed by the National Astronomical Observatory of China and Shanghai Astronomical Observatory. Data resources are supported by Chinese Astronomical Data Center (CAsDC).

SDSS-III is managed by the Astrophysical Research Consortium for the Participating Institutions of the SDSS-III Collaboration including the University of Arizona, the Brazilian Participation Group, Brookhaven National Laboratory, Carnegie Mellon University, University of Florida, the French Participation Group, the German Participation Group, Harvard University, the Instituto de Astrofisica de Canarias, the Michigan State/Notre Dame/JINA Participation Group, Johns Hopkins University, Lawrence Berkeley National Laboratory, Max Planck Institute for Astrophysics, Max Planck Institute for Extraterrestrial Physics, New Mexico State University, New York University, Ohio State University, Pennsylvania State University, University of Portsmouth, Princeton University, the Spanish Participation Group, University of Tokyo, University of Utah, Vanderbilt University, University of Virginia, University of Washington, and Yale University.

Based on observations obtained with MegaPrime/MegaCam, a joint project of CFHT and CEA/IRFU, at the Canada-France-Hawaii Telescope (CFHT) which is operated by the National Research Council (NRC) of Canada, the Institut National des Science de l'Univers of the Centre National de la Recherche Scientifique (CNRS) of France, and the University of Hawaii. This work is based in part on data products produced at Terapix available at the Canadian Astronomy Data Centre as part of the Canada-France-Hawaii Telescope Legacy Survey, a collaborative project of NRC and CNRS.

\end{document}